\documentclass[reprint,amsmath,amssymb,aps,prl,superscriptaddress,longbibliography]{revtex4-1}
\usepackage[utf8]{inputenc}
\usepackage[english]{babel}
\usepackage{amsfonts}
\usepackage{graphicx}
\usepackage{xcolor}
\usepackage{hyperref}
\usepackage[caption=false]{subfig}

\newcommand{\br}[1]{\left( #1 \right)}
\newcommand{\bra}[1]{\left< #1 \right|}
\newcommand{\ket}[1]{\left| #1 \right>}

\newcommand{\abs}[1]{\left| #1 \right|}
\newcommand{\dif}{\text{d}}
\newcommand{\kp}{{\vec{k}_\parallel}}
\renewcommand\vec{\mathbf}

\graphicspath{{figures/}}

\begin{document} 

\title{Infinite Berry curvature of Weyl Fermi arcs}

\author{Dennis Wawrzik}
\affiliation{Institute for Theoretical Solid State Physics, IFW Dresden, Helmholtzstr.\ 20, 01069 Dresden, Germany}
\author{Jhih-Shih You}
\affiliation{Department of Physics, National Taiwan Normal University, Taipei 11677, Taiwan}
\author{Jorge I. Facio}
\affiliation{Institute for Theoretical Solid State Physics, IFW Dresden, Helmholtzstr.\ 20, 01069 Dresden, Germany}
\author{Jeroen van den Brink}
\affiliation{Institute for Theoretical Solid State Physics, IFW Dresden, Helmholtzstr.\ 20, 01069 Dresden, Germany}
\affiliation{Institut für Theoretische Physik and Würzburg-Dresden Cluster of Excellence ct.qmat, Technische Universität Dresden, 01062 Dresden, Germany}
\author{Inti Sodemann}
\affiliation{Max Planck Institute for the Physics of Complex Systems, N\"othnitzer Str.\ 38, 01187 Dresden, Germany}

\date{\today}

\begin{abstract}
We show that Weyl Fermi arcs are generically accompanied by a divergence of the surface Berry curvature scaling as $1/k^2$, where $k$ is the distance to a hot-line in the surface Brillouin zone that connects the projection of Weyl nodes with opposite chirality but which is distinct from the Fermi arc itself. Such surface Berry curvature appears whenever the bulk Weyl dispersion has a velocity tilt toward the surface of interest. This divergence is reflected in a variety of Berry curvature mediated effects that are readily accessible experimentally, and in particular leads to a surface Berry curvature dipole that grows linearly with the thickness of a slab of a Weyl semimetal material in the limit of long lifetime of surface states. This implies the emergence of a gigantic contribution to the non-linear Hall effect in such devices.
\end{abstract}

\maketitle

\begin{figure}
  \centering
    \includegraphics[width=0.48\textwidth]{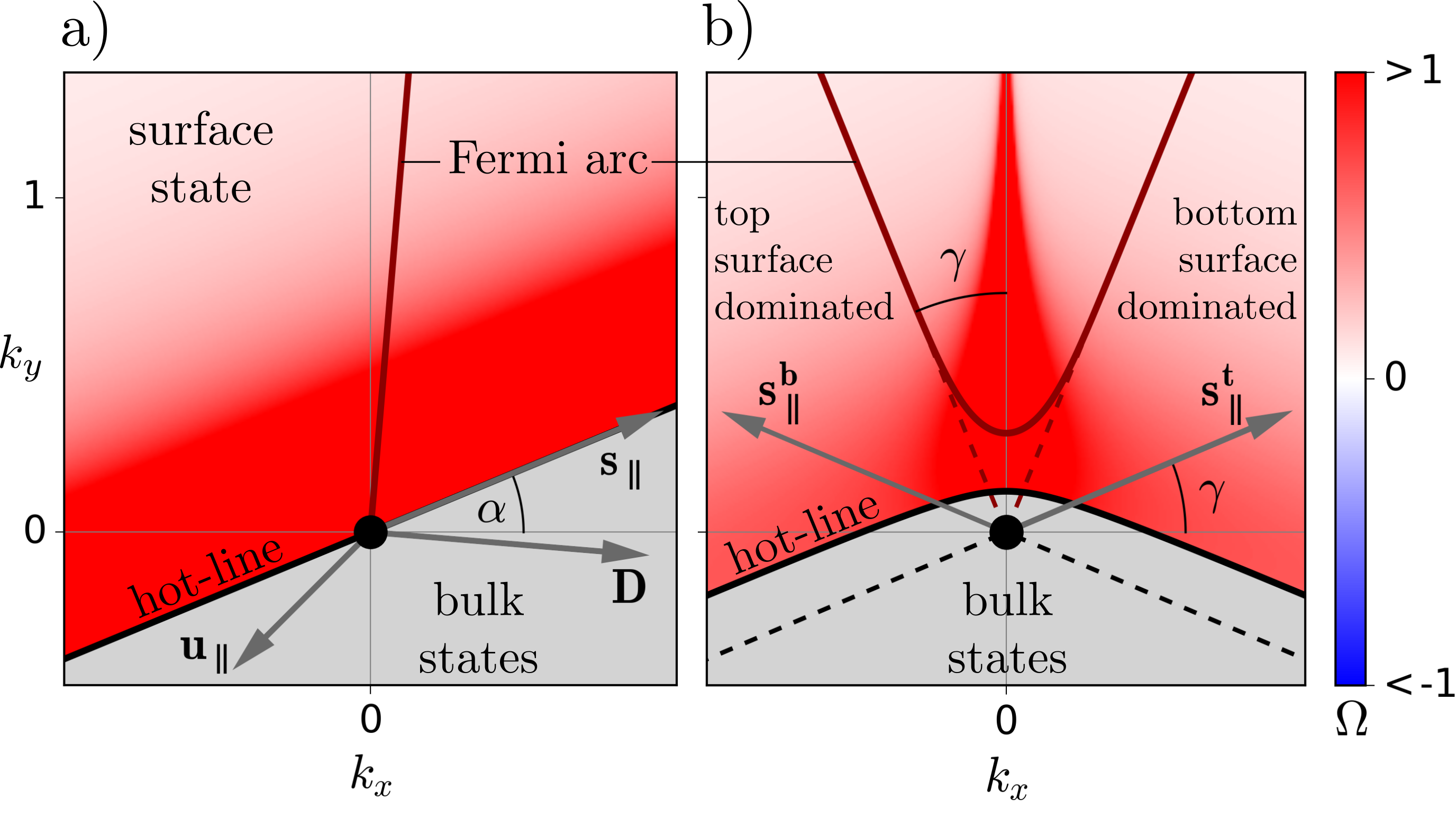}
  \caption{Hot-line (black line), Fermi arc (red line) and projected 
Weyl nodes (black dots) of a single Weyl-node surface state for a semi-inifinite (a) or finite (b) Weyl semimetal slab.  The gray area contains only surface-delocalized 3D bulk states. The red-blue color indicates the BC of the surface-localized states. In a), the in-plane component of the tilt $\vec{u}_\parallel$, the pseudo-spin polarization $\vec{s}_\parallel$ and the BC dipole $D$ are shown. (b) Top and bottom surfaces have spin polarization $\vec{s}_\parallel^{t/b}$, whereas the parameter $2\gamma$ is the opening angle between top ($k_x<0$) and bottom ($k_x>0$) arc for $\vec{u}_\parallel = 0$. Dashed lines denote the Fermi arcs and the hot-line for $L\rightarrow\infty$. The labels ``top/bottom surface dominate'' describe the surface where the  wavefunction has the most weight.
  }
  \label{fig:arc_infinite/finite}
\end{figure}

{\it Introduction.\, } The Berry curvature (BC) of electronic bands is known to deeply influence a variety of properties and phenomena~\cite{RevModPhys.82.1959}. Therefore, one of central quests of modern material design is to find mechanisms that can lead to large enhancements of BC. Typically large BC concentrations occur at points in Brillouin zone (BZ) where two bands are exactly or nearly degenerate, leading to rapid changes of the Bloch wavefunctions near such hot-spots. Classic examples of such BC hot-spots include the Weyl nodes in 3D materials~\cite{RevModPhys.90.015001,annurev-conmatphys-031016-025458} and the Dirac points in 2D graphene. In this letter, we will demonstrate a distinct mechanism by which a divergence of BC occurs not at a point, but instead over an entire line in the BZ, which we will refer to as a ``hot-line''. These hot-lines emerge in the surface BZ of Weyl semimetals~\cite{RevModPhys.90.015001,annurev-conmatphys-031016-025458} (WSMs), and they separate the 2D states that are localized at the surface from the continuum of 3D bulk states. Therefore, whereas the hot-line is distinct from the surface Fermi arc, they will meet at the momentum where the Fermi arc terminates and mixes with 3D bulk states, as illustrated in Fig.~\ref{fig:arc_infinite/finite}. The BC divergence at these hot-lines will be generically present in the WSM surface states whenever a bulk Weyl node has a velocity-tilt with a component normal to the surface of interest.

This large enhancement of BC at a hot-line is expected to manifest in a variety of contributions to physical effects in WSMs. In particular, we will demonstrate that it can lead to gigantic contribution to the BC driven non-linear Hall effect~\cite{deyo2009semiclassical,PhysRevLett.105.026805,PhysRevLett.115.216806,PhysRevB.92.235447,PhysRevLett.123.196403,Xu2018ElectricallySB,PhysRevB.98.121109,ma2019observation,kang2019nonlinear,PhysRevLett.121.246403,PhysRevB.97.041101,shvetsov2019nonlinear,PhysRevLett.123.246602}, featuring a contribution to the Berry-curvature-dipole (BCD), $D$, that grows linearly proportionally with the thickness, $L$, of a WSM slab:
\begin{align}
  D \propto L.
\end{align}
This contribution can arise even when there is no bulk Fermi surface (chemical potential is at the bulk Weyl node), so that there is nominally no 3D bulk BCD. Our findings not only enhance the understanding of the geometrical aspects of the Fermi-arcs electronic structure but also should be experimentally relevant in all BC driven phenomena in Weyl semimetals.

We will illustrate this phenomena first in an ideal model containing a single Weyl node in a semi-infinite geometry which allows a simple conceptual understanding. We will then progressively extend this model to include 
the fact that Weyl nodes always come in pairs~\cite{NIELSEN1981219} and the effects of finite thickness $L$. We will then provide quantitative estimates for WSMs TaAs and TaP based on first principles calculations.


{\it Single Weyl node model.\, } We consider a low energy model of a single Weyl cone ($\hbar=1$)
\begin{align}
\label{eq:Hamiltonian}
H = -i\vec{\nabla} \cdot \br{\chi v_F \boldsymbol{\sigma} + \vec{u} \sigma_0},
\end{align}
with chirality $\chi=\pm 1$, Fermi velocity $v_F>0$, and tilt velocity $\vec{u}$. Here $\boldsymbol{\sigma}$ is the vector of Pauli matrices and 
$\sigma_0$ is the identity. In the following we set $v_F=1$ and without 
loss of generality we assume that the surface normal $\vec{n}$ of a semi-infinite WSM residing at $z<0$ points into the $z$-direction, $\vec{n}=\hat{z}$. The wave function is therefore viewed as having non-zero amplitude only for $z\leq 0$. As discussed in Refs. \cite{PhysRevB.92.201107, witten2016three, 10.1093/ptep/ptx053, PhysRevB.97.075132, PhysRevB.100.155131}, in order to guarantee hermiticity of the Hamiltonian from Eq.~(\ref{eq:Hamiltonian}), the following boundary condition of vanishing current density normal to the surface must be imposed:
\begin{align}
\label{eq:psi_0}
\psi^\dagger \sigma_z \psi |_{z=0} = - \chi u_z\psi^\dagger \psi |_{z=0}.
\end{align}
Therefore, remarkably, the normal component of the pseudo-spin density at 
the boundary is determined by the normal component of the tilt velocity, leaving its component parallel to the boundary as a free parameter. We therefore parametrize the orientation of the pseudo-spin density at the boundary by a unit vector $\vec{s}$:
\begin{align}
\label{eq:s}
\vec{s} = \left(\sqrt{1-u_z^2} \cos (\alpha), \sqrt{1-u_z^2} \sin(\alpha), - \chi u_z \right).
\end{align}
Within our continuum model of semi-infinite space, $\alpha$ is an undetermined parameter. However, $\alpha$ is understood to be uniquely fixed in an underlying microscopic description. One way to do this is to solve a continuum model in the entire space that includes an explicit description of the decay of the wave-function into vacuum, as we illustrate in the Suppl. Materials~\cite{Supp} in Sec.\ B. Moreover, although $\alpha$ is unique in any microscopic Hamiltonian defined for all $z$, it can depend on microscopic details of the boundary~\cite{10.1093/ptep/ptx053} and on band-bending effects~\cite{PhysRevB.92.201107}. In writing Eq.~(\ref{eq:s}) we are assuming that the system is a type-I WSM~\cite{soluyanov2015type} with $|\vec{u}| < v_F$.

Wave functions localized at the boundary have the form $\Psi\br{\kp,z} = 
e^{\lambda z+i\kp \cdot \vec{r}_\parallel} \psi$ with Re$(\lambda) > 0$, and $\vec{r}_\parallel$, $\kp$ denoting the coordinates and momentum parallel to the boundary. Their dispersion is found to be (see Suppl. Materials \cite{Supp}, Sec.\ A)
\begin{align}
\label{eq:energy_inf}
E_\kp  = \kp \cdot \br{\chi\vec{s}+\vec{u}},
\end{align}
and
\begin{align}
\label{eq:lambda}
\lambda = -\frac{\kp\cdot \br{\vec{n}\times\vec{s} + i \chi u_z \vec{s}}}{1-u_z^2}.
\end{align}
The Fermi arc at $E_\kp=0$ is therefore a straight ray pointing in the direction $\vec{n} \times \br{\chi\vec{s}+\vec{u}}$ and starting from the 
projection of the Weyl node onto the surface of interest as illustrated in Fig.~\ref{fig:arc_infinite/finite} (a). The surface BC is given by 
$\Omega_z=\partial_{k_x}A_y-\partial_{k_y}A_x$ with Berry connection
\begin{equation}
\vec{A}\br{\kp} = -i \int_{-\infty}^{0} \dif z ~\Psi^\dagger\br{\kp,z} \vec{\nabla}_{\kp}\Psi\br{\kp,z},
\end{equation}
which yields:
\begin{align}
\vec{\Omega}\br{\kp} = -\frac{\chi u_z \br{1-u_z^2}}{2 \br{\kp\cdot\br{\vec{n}\times\vec{s}}}^2}\vec{n},
\end{align}
where $|\vec{n}\times\vec{s}|^2 = 1-u_z^2$ and is fully determined by the tilt $\vec{u}$, the surface normal $\vec{n}$, chirality $\chi$, and boundary spin polarization direction $\vec{s}$. The equation above summarizes one of our central findings: The BC of the surface diverges quadratically as the surface momentum approaches the hot-line defined by $\kp\cdot\br{\vec{n}\times\vec{s}}=0$ which separates the states localized at the 
surface, that have Re$(\lambda)>0$ in Eq.~(\ref{eq:lambda}), from those that mix with the 3D bulk (see Fig.~\ref{fig:arc_infinite/finite} (a)). We emphasize that this divergence occurs over an entire line in the surface momentum plane (the surface BZ in a lattice model) and not just on the single point. Notice also that this hot-line is distinct from the Fermi arc, but the Fermi arc always terminates at a point in this line independent of the Fermi energy. Furthermore, the BC is proportional to $u_z$ and vanishes if the Weyl cone is not tilted toward the surface. Since $u_z = \vec{u}\cdot\vec{n}$ the surface state has the same BC on the top and bottom surface with $\vec{n} \rightarrow -\vec{n}$. 
Divergences of the BC typically encode a rapid change of the intra-unit cell Bloch wavefunction as a function of the crystal momentum. Although the commonly known BC hot-spots around a bulk Weyl node originate from 
the rapid change of the wavefunction pseudospin as  the node is crossed staying within the lower branch, the divergence of BC at the surface hot-line orginates from a different underlying mechanism. The BC hot-line does 
not originate from a rapid change of the pseudo-spin, whose orientation remains essentially constant upon approaching the line, but rather from an 
abrupt change of the localization of the wavefunction along the direction 
orthogonal to the surface. This is why it coincides with the line that separates bulk from surface-localized states.
In experiments this BC can be probed via the Berry curvature dipole $\vec{D}$ \cite{PhysRevLett.115.216806, PhysRevB.92.235447}. This quantity describes the first order moment of the BC over occupied states near Fermi surface and is defined in two dimensions as \cite{PhysRevLett.115.216806}:
\begin{equation}
\label{eq:BCD}
\vec{D} = -\int \frac{\dif^2\kp}{\br{2\pi}^2} ~ \Omega_z ~ \vec{\nabla}_{\kp} n_f,
\end{equation}
where $n_f$ is the Fermi-Dirac distribution. Since the BC diverges at the 
end of the Fermi arc, the Fermi arc contribution to BCD also diverges. It 
is instructive to consider the regularization of this divergence by introducing a cutoff at distance $k_c$ from the hot-line in momentum space. We 
will discuss how to obtain $k_c$ later in a finite slab geometry. For a single Weyl cone we then find the following Fermi arc contribution to the surface BCD at zero temperature
\begin{align}
\label{eq:BCD_cutoff}
\vec{D} = -\frac{\chi u_z}{8\pi^2 k_c}\sqrt{1-u_z^2}\frac{\vec{s} 
+ \chi\vec{u}}{1+\chi \vec{s}\cdot\vec{u}},
\end{align}
which thus scales as $D\propto u_z/k_c$ and is perpendicular to the Fermi 
arc as depicted in Fig.~\ref{fig:arc_infinite/finite} (a). Although we have taken zero temperature, one can verify that the Fermi arc contribution to the BCD remains infinite at finite temperature and it is independent of the chemical potential.

\begin{figure}
\centering
  \includegraphics[width=0.48\textwidth]{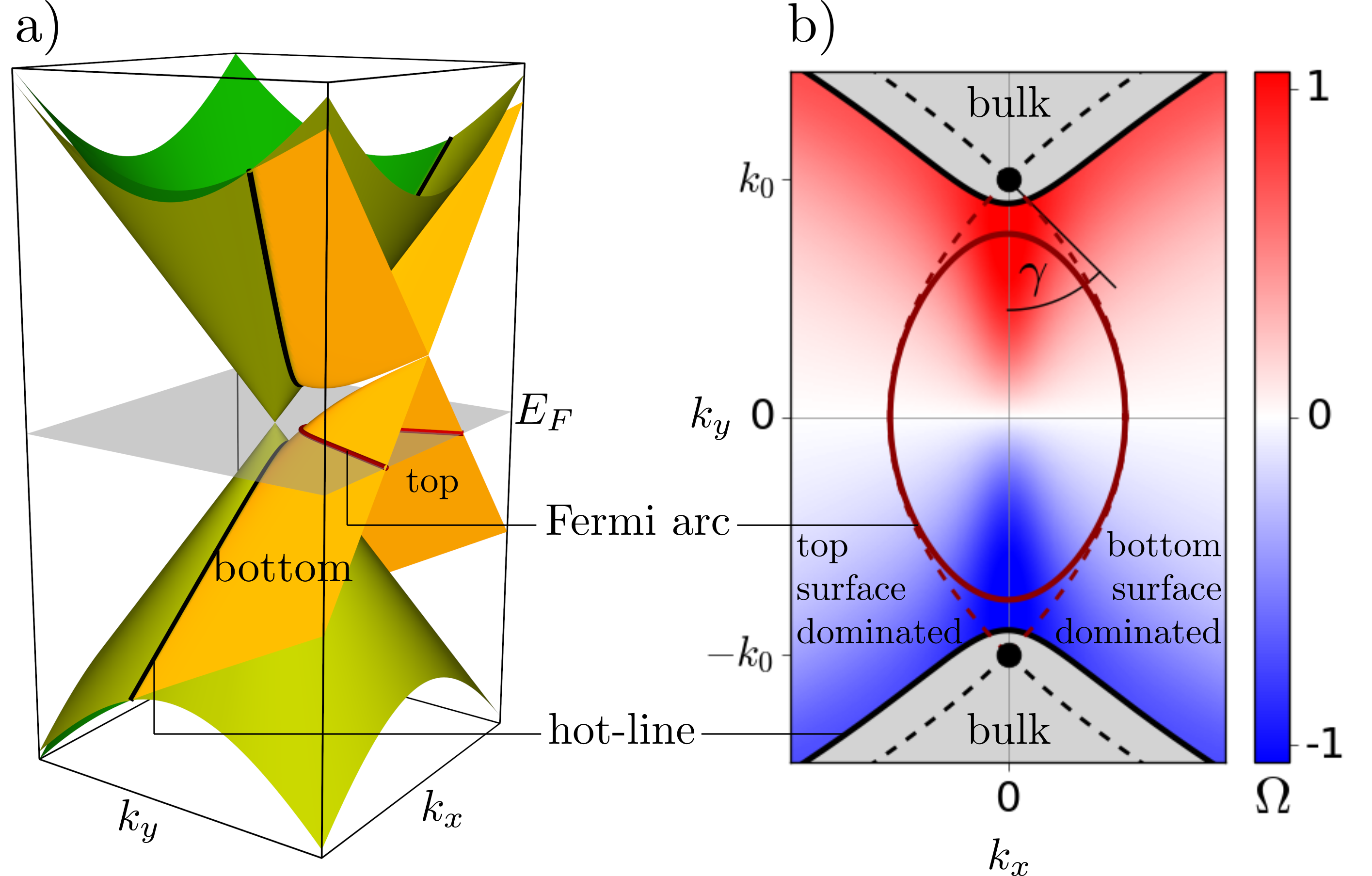}
  \caption{(a) Energy dispersion for a single Weyl node in a slab of width $L$. The orange bands are located on the surface, whereas the green ones correspond to bulk states, separated by the black hot-line. The red Fermi arc, here for chemical potential $\mu=0$, does not end at the Weyl node in a finite system. (b) Hybridized Fermi arcs (solid red) for a Weyl pair for $L=4/k_0$ and the Berry curvature $\Omega$. The dashed lines mark the Fermi arcs and hot-lines for $L\rightarrow\infty$. The hot-line (solid black) separates the states localized at the surface (color region) from the region containing only bulk states (gray region). The surface where the wavefunction has the most of weight is labeled as in Fig.\ 1(b).}
  \label{fig:ES_finite}
\end{figure}

{\it Weyl pair model.\, } Fermi arcs on crystal surfaces always connect projections of two Weyl nodes with opposite chirality. In order to get a low-energy Hamiltonian for a pair of Weyl cones \cite{PhysRevLett.121.246403,PhysRevB.100.155131,PhysRevLett.123.246602} we modify the Hamiltonian in Eq.~(\ref{eq:Hamiltonian}) by substituting 
\begin{align}
\label{eq:ky_sub}
k_y \rightarrow \tilde{k}_y = \frac{k_0^2-k_y^2}{2k_0}.
\end{align}
The two Weyl nodes with chirality $\pm \chi$ are located at $k_y=\mp k_0$ ($k_0 > 0$) and $k_x=k_z=0$ and are related by a mirror symmetry $M_y$. This symmetry acts only in the spatial part but not the pseudospin $\boldsymbol{\sigma}$, and thus it is not broken by the boundary condition in Eq.~(\ref{eq:psi_0}). For simplicity, we take here $\vec{u} = u_z \vec{n}$. Again, if we perform an analogous analysis as in the single Weyl 
node we are led to boundary conditions parametrized by the vector $\vec{s}$ from Eq.~(\ref{eq:s}). The Fermi arc becomes a parabola between the nodes as illustrated in Fig.~\ref{fig:ES_finite} (b) where the free parameter $\alpha = \gamma$ is the angle between $k_y$-axis and the 
Fermi arc at the node. The arc connects the Weyl nodes for $|\alpha| < \pi/2$, in a lattice model the nodes would also be connected for larger values of $\alpha$ via the periodic boundaries of the lattice BZ. We have verified that an alternative model with an explicit interface between vacuum and the WSM recovers the results of this model for the special case of $\alpha=0$ (see SM~\cite{Supp}, Sec.\ B).

In this semi-infinite system the surface BC is again diverging at the ends of the Fermi arc and the hot-line becomes a parabola, and it is given by
\begin{align}
\Omega\br{\kp} = -\frac{\chi u_z k_y}{2 k_0 \br{k_x \sin(\alpha)-\tilde{k}_y \cos(\alpha)}^2}.
\end{align}
Notice that as a consequence of the $M_y$ mirror symmetry the BC changes sign at $k_y=0$ and the BCD only has non-zero $D_y$ component~\cite{PhysRevLett.115.216806}. Since most of the BC of the Fermi arc is located close to the nodes, the size of the BCD of a Weyl pair is approximately twice the $D_y$ component of the BCD of a single node.

{\it Weyl nodes in finite slabs.\, } As we have discussed there is a remarkable divergence of the surface BC and the surface BCD of ideal Fermi arcs in semi-infinite geometries. In real materials a variety of physical effects can regularize such divergence such as the finite lifetime of surface states induced by disorder or phonons. Here we will focus on the limit in which the lifetime associated with the decay of surface quasiparticles into the bulk is much longer than the intra-surface scattering time. In this limit the regularization of these effects arises from the finite thickness of the material. To do so we restrict the WSM to reside within $|z| \leq L/2$. As detailed in the SM~\cite{Supp}, Sec.\ C, hermiticity now requires $\chi s_z - u_z$ to be the same on both surfaces. Nevertheless, the parallel component $\vec{s}_\parallel$ is again not fixed and can have a different angle on top and bottom surface. We define these two free parameters to be the angle $\pi-2\gamma$ between the top and bottom spin polarization with $|\gamma| < \pi/2$, and the angle $\theta$ between $\vec{s}_\parallel^t$ for $\gamma = 0$ and positive $k_x$-axis. Without loss of generality we can set $\theta = 0$, i.e. for $\vec{u}_\parallel=0$ the Fermi arcs are located around the $k_y$-axis with angle $\pm\gamma$ as shown in Fig.~\ref{fig:arc_infinite/finite} (b).

The energy dispersion of the surface states can be conveniently parametrized in terms of the function $\epsilon(\kp)$, as $E\br{\kp} = \chi 
\sqrt{1-u_z^2} \epsilon(\kp) + \kp\cdot\vec{u}$, which in turn is defined 
implicitly as a solution of the following transcendental equation:
\begin{align}
\label{eq:energy_tr}
\sqrt{\kp^2-\epsilon^2} = \frac{k_y - \epsilon \sin(\gamma)}{\cos(\gamma)} \tanh\br{L \sqrt{\frac{\kp^2-\epsilon^2}{1-u_z^2}}}.
\end{align}
The resulting energy dispersion is illustrated in Fig.~\ref{fig:ES_finite} (a). The decay length of these solutions into the bulk is given by Re$(\lambda)\propto \sqrt{\kp^2-\br{\epsilon(\kp)}^2}$. Therefore the surface states of our interest correspond to the two solutions with $\br{\epsilon(\kp)}^2 < \kp^2$. In the $L\rightarrow \infty$ limit this equation recovers the energy dispersion from Eq.~(\ref{eq:energy_inf}) for each surface. Fig.~\ref{fig:ES_finite} shows that for finite $L$ a gap opens at the intersection of the two orange surface bands, which decreases exponentially in system size and distance from the Weyl node. This opening of a gap gives rise to the fact that neither the Fermi arc nor the hot-line touches the Weyl point anymore but hybridizes with the arc/line from the opposite surface for finite $L$. The Fermi arc and the hot line will, however, touch the surface projection of the Weyl node for $L\rightarrow\infty$ (see Fig.\ \ref{fig:arc_infinite/finite} (b) and Fig.\ 1 in SM~\cite{Supp} for an illustration). A first order Taylor expansion of Eq.\ (\ref{eq:energy_tr}) around $\epsilon^2=\kp^2$ determines the BC hot-line depicted as a solid black line in Figs.\ 
\ref{fig:arc_infinite/finite} (b) and \ref{fig:ES_finite} where the surface state becomes a bulk state. The minimal distance of the BC hot-line to the Weyl node is given at $k_x=0$ and $k_y = k_c$ with 
\begin{align}
k_c = \frac{\sqrt{1-u_z^2}}{L \cos (\gamma)}(1-\sin(\gamma)),
\end{align}
which serves as a momentum cutoff for the BCD, as discussed in the previous section.

This leads to the remarkable conclusion that, ideally, the Fermi arc contribution to the BCD increases with the thickness $L$ of the crystal slab. Therefore, since our model of a single Weyl node contains as its only length scale the slab thickness $L$, and the BCD has units of length in 2D~\cite{PhysRevLett.115.216806}, this model predicts a BCD that is exactly linearly proportional to $L$, without the need of any explicit cutoff. Aditionally, Fig.~\ref{fig:ES_finite} (b) shows the direct numerical calculation of the BC for our model of a pair of Weyl nodes from Eq.~(\ref{eq:ky_sub}) in a finite slab geometry. As we see there are some notable modifications brought in by the pair of Weyl cones in a finite size. In particular the BC is highly peaked around $k_x=0$, i.e. where the surface state changes from top to bottom surface.

{\it Experimental signatures.\,} WSMs exist only in materials breaking either spatial inversion or time-reversal (TR) symmetry. Our results can be 
applied generally to materials that break either of these symmetries by considering additional symmetry related copies of the elementary models of 
Weyl cones that we have studied, and thus Weyl Fermi arcs generically display the infinite Berry curvature at hot-lines that we have described. In 
the special case in which the material is time reversal invariant and the 
non-linear Hall effect is the leading Hall-like reponse, the surface BCD will remain non-zero after adding symmetry related copies of the models. More precisely, two Weyl node pairs that are related by TR will have additive contributions to the BCD.

Candidate materials for observing the giant non-linear Hall effect resulting from this Fermi arc contribution to the BCD include, in principle, all 3D type-I WSMs with broken inversion symmetry. For example, WSMs TaAs and 
TaP~\cite{PhysRevB.97.041101} are TR symmetric and well suited for experiments. We performed density-functional (DFT) calculations for these two compounds using a generalized gradient approximation~\cite{PhysRevLett.77.3865} as implemented in the FPLO code~\cite{PhysRevB.59.1743}. As in previous reports~\cite{PhysRevX.5.011029,lv2015observation}, we find a set of 
four pairs of Weyl nodes in the $k_z = 0$ plane and a second set of eight Weyl nodes away from such plane. In the following, we will focus in the latter set, named $W_2$, which is of type I in TaAs and TaP, and placed 
closer to the Fermi energy. Detailed values for the fit of the DFT data to our model can be found in the SM~\cite{Supp} in Sec.\ D. WSMs of the TaAs family belong to the point group $C_{4v}$, i.e. these materials have a polar axis in $z$-direction allowing a bulk BCD~\cite{PhysRevLett.121.246403,PhysRevLett.123.246602,PhysRevB.97.041101}.

In order to have a non-vanishing net BCD, the maximal symmetry of the finite system including its surfaces should be a mirror line~\cite{PhysRevLett.115.216806}. Such mirror line forces the direction of the BCD to be perpendicular to it. Furthermore, symmetries such as bulk mirror planes parallel to the surface must be avoided since they project Weyl nodes of opposite chirality onto the same point in the surface Brillouin zone. It is interesting, in particular, to consider materials that are nominally inversion invariant in the bulk, but the 3D inversion of the slab is broken by a surface effect, such as the asymmetry between top and bottom surfaces that one would have when the slab resides in a substrate. In these cases, the non-linear Hall effect will be dominated by the surface states allowing one to directly probe the divergence of the surface Berry curvature of the Weyl Fermi arcs. For WSMs of the TaAs family the crystal symmetries allow all cones to be tilted perpendicular to the 4-fold rotation axis. Due to this rotation axis the (001) surface does not exhibit a surface BCD, as well as a surface parallel to one of the mirrors belonging to this axis. Nevertheless, a suitable surface which keeps a single mirror line $M_x$ and has proper tilted cones is, for instance, (011). Depending on $\alpha$, the BCD of a single W$_2$ Weyl pair can reach up to $D_x = 0.042L$ in TaAs and $D_x=0.041L$ in TaP on this surface.

The BC divergence at the hot-line should have interesting consequences in 
many phenomena~\cite{RevModPhys.82.1959}. In addition to the non-linear transport signatures, the BCD $\vec{D}$ also induces an effective orbital magnetization $\vec{M} \propto \vec{D}\cdot\vec{E} \hat{e}_z$ in the presence of an electric field $E$ which can be measured using Kerr rotation or circular dichroism~\cite{PhysRevLett.123.036806}. A further e!xperimental probe is the second order response to an applied thermal gradient $\nabla T$~\cite{PhysRevB.99.201410}. The nonlinear anomalous Nernst effect is a current perpendicular to the temperature gradient and proportional to $(\nabla T)^2$, which at low temperatures is linear in the BCD~\cite{PhysRevResearch.2.032066}. Other interesting potential manifestations of this divergence could be appear in the surface contributions to the orbital magnetic moment of materials with bulk Weyl nodes \cite{PhysRevB.59.14915,PhysRevB.98.155145,PhysRevB.101.201402}, and also in the surface contributions in the non-linear version of the chiral anomaly \cite{PhysRevB.103.045105}.

{\it Concluding remarks. \,} We have demonstrated the existence of a generic divergence of the Berry curvature at a hot-line in the surface of Weyl semimetals, which is distinct from the Fermi arc itself. This divergence scales as the inverse square distance in momentum space to the hot-line. Such divergence leads to a variety of strong Berry phase driven effects. We have focused on the implications of this divergence on the Berry curvature dipole, which controls the non-linear Hall effect and found the Fermi arc contribution to the BCD itself grows linearly with the width $L$ of the Weyl semimetal slab, suggesting a gigantic enhancement of the non-linear Hall effect in such devices even when the bulk has nominally zero BCD because of the chemical potential is at the bulk Weyl node. Our description can be viewed as valid in the limit in which the lifetime for decay of surface quasiparticles into the bulk is much longer than the intra-surface scattering time. In general disorder or surface imperfections~\cite{PhysRevB.93.235127,PhysRevB.96.201401,PhysRevB.97.235108,PhysRevX.8.031076,PhysRevX.8.031076,PhysRevB.99.155404,PhysRevB.100.165422,PhysRevB.100.195117,du2019disorder,Isobeeaay2497} and phonon scattering~\cite{PhysRevX.9.021053,PhysRevX.9.031036,PhysRevB.100.220301,PhysRevB.101.085202,PhysRevB.102.125104,PhysRevLett.125.146402} can degrade such surface life-time. However, our description should be a good approximation in high quality surfaces where the surface crystal momentum conservation enhances the lifetime of quasiparticles, and at low temperatures where the phonon contribution to scattering becomes small in comparison with elastic impurity scattering. This limit should be experimentally accesible in high-quality surfaces, such as those where very sharp and long-lived surface Fermi arc states have been imaged by quasi-particle interference experiments~\cite{Inoue1184}, or those in which highly energy-momentum resolved Fermi arc states have been imaged by ARPES as reported e.g.\ in Refs.\ \cite{xu2015discovery,PhysRevB.94.121112}.
Another future direction is understanding the fate of our hot-lines in other topological semimetals~\cite{Bradlynaaf5037,PhysRevB.98.155145}, and, more broadly speaking, their role in the bulk-boundary correspondence of gap-less topological phases, which remains to be fully elucidated.

{\it Acknowledgements. \,} We thank Ion Cosma Fulga for insightful discussions and Ulrike Nitzsche for technical assistance. J.I.F.~acknowledges support from the Alexander von Humboldt Foundation and D.W.~thanks the IFW excellence programme. J.v.d.B.~acknowledges financial support from the German Research Foundation (Deutsche Forschungsgemeinschaft, DFG) via SFB 1143 Project No.~A5 and under Germany's Excellence Strategy through the W\"urzburg-Dresden Cluster of Excellence on Complexity and Topology in Quantum Matter ct.qmat (EXC 2147, Project No.~390858490). J.-S. Y. is supported by the Ministry of Science and Technology, Taiwan (Grant No. MOST 110-2112-M-003 -008 -MY3) and National Center for Theoretical Sciences in Taiwan.

\newpage
\section*{Supplementary information}
\subsection{A\quad Surface wave function for semi-infinite system}
\label{sec:semi-infinite}
Here we give the full surface wave function of a semi-infinite Weyl semimetal (WSM) with a tilted cone. We start from the low energy Hamiltonian
\begin{align}
	\label{eq:Hamiltonian}
	H = -i\vec{\nabla} \cdot \br{\chi \boldsymbol{\sigma} + \vec{u} \sigma_0}
\end{align}
and restrict it to $z \leq 0$ such that we get a surface with normal $\vec{n} = \hat{z}$. This makes it convenient to split every vector in a parallel and perpendicular part, e.g.\ $\vec{u} = \vec{u}_\parallel + u_z \vec{n}$ with $\vec{u}_\parallel \perp \vec{n}$. The Hamiltonian is translational invariant parallel to the surface which allows us to replace $-i\vec{\nabla} \rightarrow \kp -i\vec{n}\partial_z$. By partial integration one can show that hermiticity $\left< \psi_1,H \psi_2\right> = \left<H \psi_1, \psi_2\right>$ of the Hamiltonian requires the boundary term at $z=0$ to vanish \cite{PhysRevB.92.201107, witten2016three, 10.1093/ptep/ptx053, PhysRevB.97.075132, PhysRevB.100.155131}, i.e. 
\begin{align}
	\label{eq:psi_0}
	\psi_1^\dagger \br{\chi \sigma_z + u_z}\psi_2 |_{z=0} = 0
\end{align}
for all $\psi_1,\psi_2$. Here we look at the special case where $\psi_1=\psi_2 \equiv \psi$, $\abs{\psi}^2=1$, are $z$-independent spinors so that the expectation value $\vec{s}$ of the pseudo-spin for $\psi \equiv \psi(\alpha)$ can be written as 
\begin{align}
\label{eq:s}
	\vec{s} = \left(\sqrt{1-u_z^2} \cos (\alpha), \sqrt{1-u_z^2} \sin(\alpha), - \chi u_z \right)
\end{align}
with free real parameter $\alpha$ labeling all possible boundary conditions. For the surface wave function $\Psi$ we make the ansatz 
\begin{align}
	\label{eq:ansatz}
	\Psi\br{\vec{k_\parallel},z} = c\br{\kp} e^{i\kp\cdot\vec{r}_\parallel + \lambda\br{\kp} z} \psi\br{\alpha}
\end{align}
where $c(\kp)$ is the (real) normalization constant. Since $\Psi \propto \psi(\alpha)$ for all $z$ the boundary condition Eq.\ (\ref{eq:psi_0}) always is fulfilled. In order to obtain a surface state the function $\lambda(\kp)$ must have a non-vanishing real part. More precisely, normalization $1 = \int_{-\infty}^0 \dif z \abs{\Psi}^2$ gives $c^2 = 2$Re$(\lambda)$ and requires Re$(\lambda) > 0$. 
Applying the gradient of the Hamiltonian in Eq.\ (\ref{eq:Hamiltonian}) to wave function $\Psi$ yields the effective Hamiltonian
\begin{align}
	H_{\text{eff}} = \kp\cdot \br{\chi\boldsymbol{\sigma} + \vec{u}_\parallel\sigma_0} -i\lambda\br{\chi\sigma_z+u_z\sigma_0}.
\end{align}
Note that we are not dealing with non-hermitian Hamiltonians since $\br{\chi\sigma_z+u_z\sigma_0}\Psi = 0$ due to the boundary condition. Thus, the energy $H\Psi = H_\text{eff}\Psi = E \Psi$ becomes
\begin{align}
	\label{eq:energy_inf}
	E(\kp)  = \kp \cdot \br{\chi\vec{s}+\vec{u}}.
\end{align}
The function $\lambda(\kp)$ can be obtained by solving $-\chi u_zE(\kp) = \bra{\psi} H_\text{eff} \sigma_z\ket{\psi}$ for $\lambda$. This yields 
\begin{align}
\label{eq:lambda}
	\lambda = -\frac{\kp\cdot \br{\vec{n}\times\vec{s} + i \chi u_z \vec{s}}}{1-u_z^2}.
\end{align}
Finally, from $\psi^\dagger \boldsymbol{\sigma} \psi = \vec{s}$ we find
\begin{align}
	\label{eq:psi_inf}
	\psi\br{\alpha} = \frac{1}{\sqrt{2}}
	\begin{pmatrix}
		e^{-i\alpha/2} \sqrt{1-\chi u_z} \\
		e^{+i\alpha/2} \sqrt{1+\chi u_z}
	\end{pmatrix}.
\end{align}
$~$

\subsection{B\quad Explicit interface between vacuum and WSM}
It is possible to determine the free parameter $\alpha$ if we consider a WSM at $z \leq0 $ and also the insulating phase at $z>0$. This can be achieved by modifying the Hamiltonian from Eq. (\ref{eq:Hamiltonian}) with $\vec{u}=u_z \hat{z}$ and $\chi=1$ by replacing $k_y$ with
\begin{align}
\tilde{k}_y(z) = 
\begin{cases}
\frac{+k_0^2-k_y^2}{2k_0}, & z\leq 0\\
\frac{-\Delta^2-k_y^2}{2\Delta}, & z>0
\end{cases},
\end{align}
where $2k_0$ is the distance between the Weyl nodes and $\Delta>0$ the gap of the insulator. On both sides the wave function has to decay exponentially, i.e. we have to fulfill the conditions (I) Re$(\lambda)> 0$ for the WSM and (II) Re$(\lambda)< 0$ on the gapped side for all $\kp$. From Eqs.\ (\ref{eq:s}) and (\ref{eq:lambda}) we find
\begin{align}
\text{Re}(\lambda) \propto -\sin(\alpha) k_x  + \cos(\alpha) \tilde{k}_y(z).
\end{align}
Thus, condition (II) is only satisfied for all $\kp$ if $\sin(\alpha)=0$ and $\cos(\alpha)>0$, i.e. $\alpha = 0$. Condition (I) then requires the surface state to be located at $|k_y|<k_0$ and the Fermi arc becomes a straight line between the Weyl nodes. This also holds for the perfect vacuum $\Delta\rightarrow\infty$.

\subsection{C\quad Wave function for finite system}
\begin{figure}
\label{fig:FA_HL}
  \begin{minipage}{0.238\textwidth}
    \includegraphics[width=\textwidth]{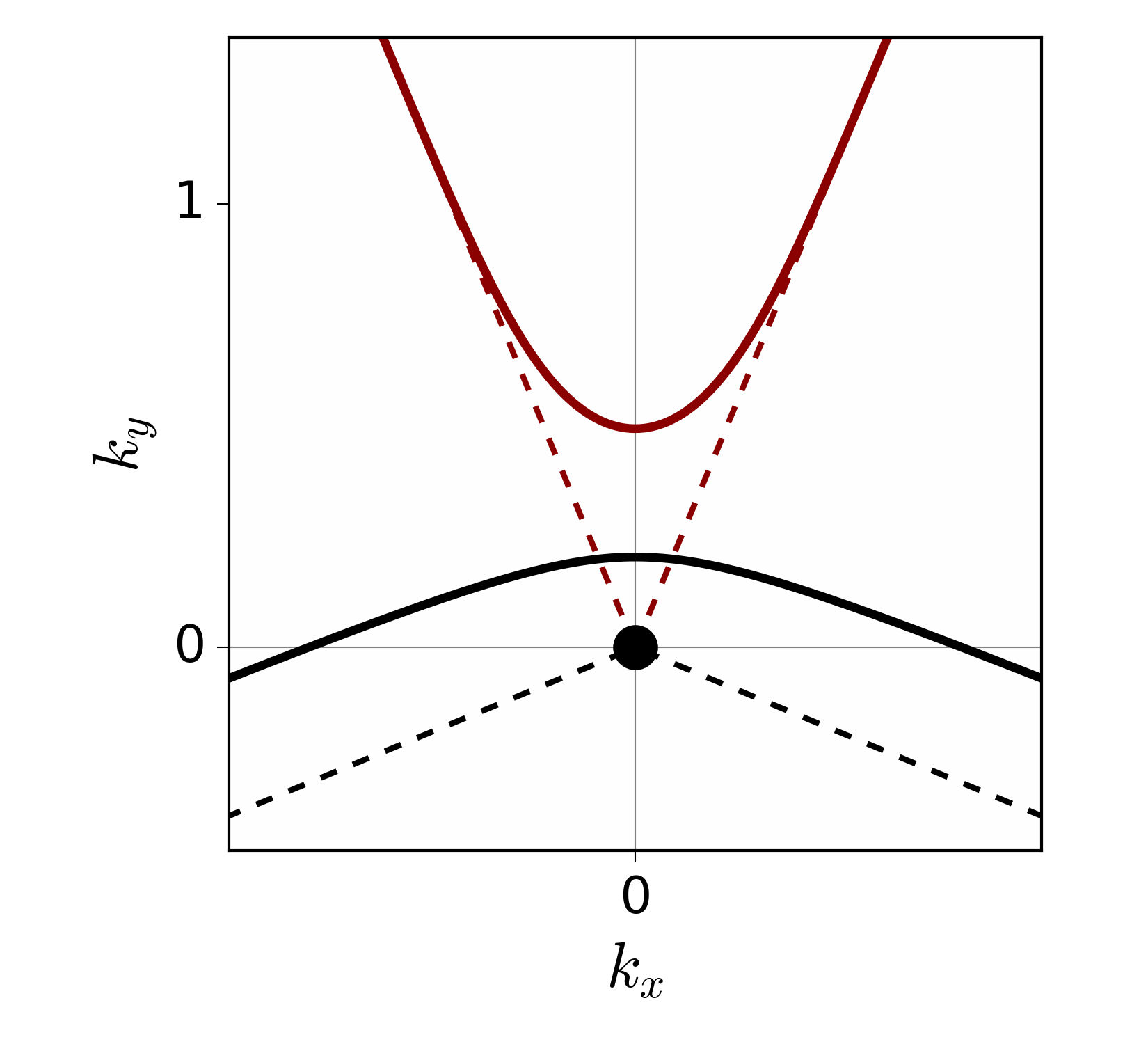}
  \end{minipage}
  \begin{minipage}{0.238\textwidth}
    \includegraphics[width=\textwidth]{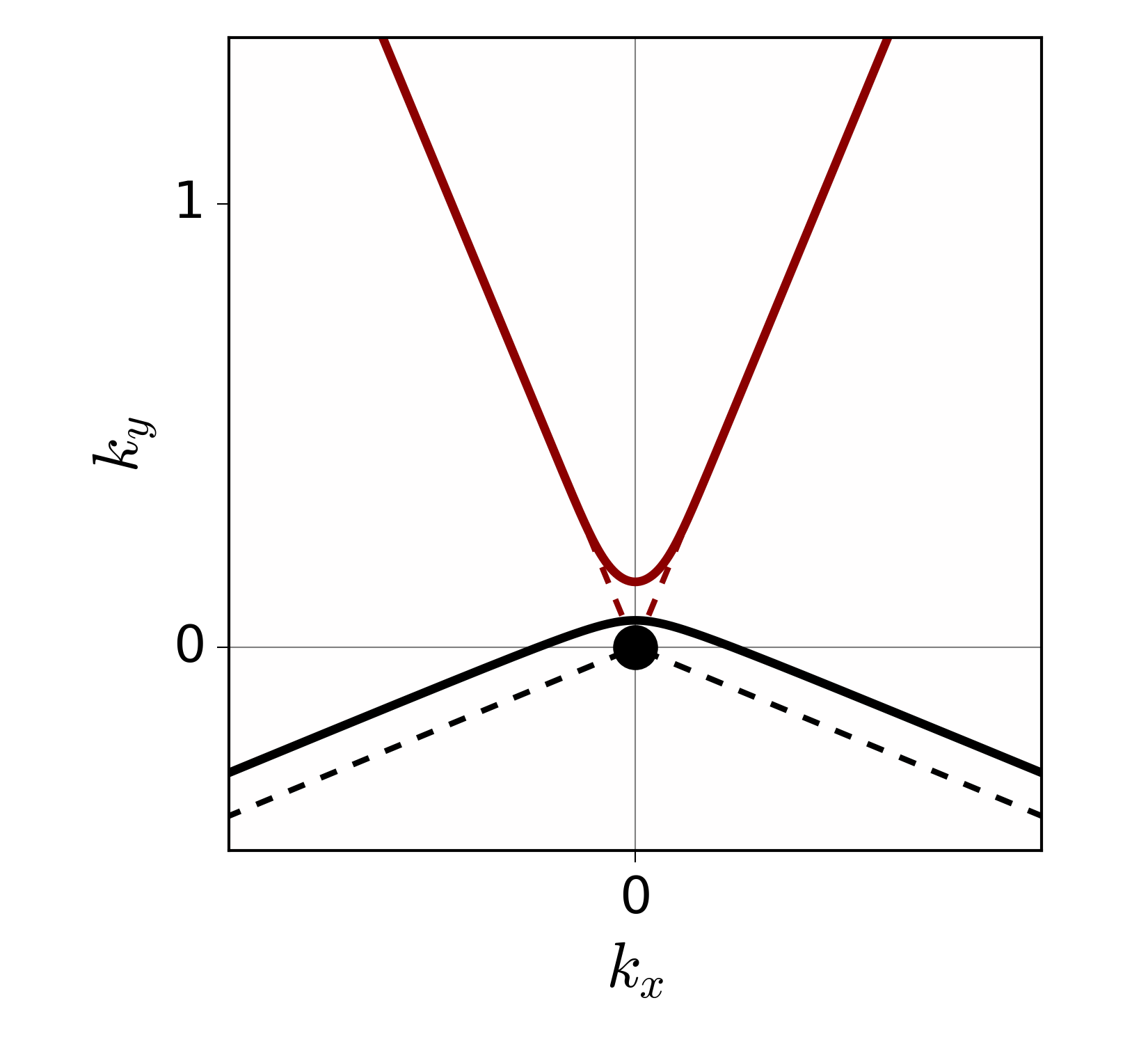}
  \end{minipage}
  \caption{Fermi arc and hot-line in a finite slab with width $L=3$ (left) and $L=10$ (right). The dashed lines mark the Fermi arc and hot-line for $L\rightarrow\infty$.}
\end{figure}

For the finite system we take the same Hamiltonian and ansatz as in the semi-infinite case (Eqs.\ (\ref{eq:Hamiltonian}) and (\ref{eq:ansatz})), restrict them to $-L/2 \leq z \leq L/2$, and solve Schr\"odinger's equation for $\lambda$ first. This results in
\begin{align}
	\lambda_s &= \frac{s\sqrt{k_\parallel^2-\epsilon^2} - i u_z \epsilon}{\sqrt{1-u_z^2}}\\
	\psi_s &= \frac{1}{\sqrt{2}}
	\begin{pmatrix}
		\frac{\epsilon - i s \sqrt{k_\parallel^2-\epsilon^2}}{k_x + i k_y} \sqrt{1-\chi u_z} \\
		\sqrt{1+\chi u_z}
	\end{pmatrix}
\end{align}
with $\epsilon(\kp)$ defined via the energy $E(\kp)$ of the surface state as $\chi\sqrt{1-u_z^2}\epsilon(\kp) = E(\kp) - \kp\cdot\vec{u}$ and $s=\pm 1$. Note that this is consistent with the results of the semi-infinite system. The full wave function is a superposition of both solutions of $\lambda$, i.e.
\begin{align}
	\Psi\br{\kp,z} = \frac{e^{i\kp\cdot\vec{r}_\parallel}}{\sqrt{2}}\sum_{s=\pm 1} s c_s e^{\lambda_s z} \psi_s
\end{align}
The condition for hermiticity now is
\begin{align}
\label{eq:psi_L}
	\psi_1^\dagger \br{\chi \sigma_z + u_z}\psi_2 |_{z=-L/2} = \psi_1^\dagger \br{\chi \sigma_z + u_z}\psi_2 |_{z=+L/2}.
\end{align}
Since our wave function will in general have different weights $\left|\Psi(z=+L/2)\right|^2 \neq \left|\Psi(z=-L/2)\right|^2$ on both surfaces, Eq.\ (\ref{eq:psi_L}) can only be satisfied if both sides of the equation are equal to zero. This leads to the same pseudo-spin polarization and wave function at the surface as Eqs.\ (\ref{eq:s}) and (\ref{eq:psi_inf}), but the free parameter $\alpha_{t/b}$ can be different on top and bottom surface. By using these boundary conditions we find an equation for energy $\epsilon(\kp)$:
\begin{align}
\label{eq:E_finite}
	\tanh\br{L \sqrt{\frac{k_\parallel^2-\epsilon^2}{1-u_z^2}}} = \frac{\sqrt{k_\parallel^2-\epsilon^2}\cos(\gamma)}{k_y\cos(\theta) - k_x \sin(\theta) - \epsilon \sin(\gamma)}
\end{align}
where $\alpha_t = \theta + \gamma$ and $\alpha_b -\pi = \theta - \gamma$. For $\theta = 0$ normalization yields
\begin{widetext}
\begin{align}
	c_s^2 &=  c_0 \frac{\br{\br{(k_x+ik_y)\sin(\gamma)-i\epsilon} \br{i\epsilon - s \sqrt{k_\parallel^2-\epsilon^2}}-k_x(k_x+ik_y)}}{L \epsilon c_0^2 - \sqrt{1-u_z^2}\cos(\gamma)\br{k_y \epsilon - k_\parallel^2\sin(\gamma)}}\\
	c_0^2&= \br{\epsilon - k_y\sin(\gamma)}^2 - \br{k_x \cos(\gamma)}^2.
\end{align}
\end{widetext}
Fig.\ (3) shows the resulting Fermi arc and hotline in dependence of system size $L$. To compute the the BC, we obtained the wave function by numerically solving Eq.\ (\ref{eq:E_finite}) and taking a finite symmetric difference quotient for the derivatives.

\subsection{D\quad Berry curvature dipole for $\text{TaAs}$ and $\text{TaP}$}
In order to acess the parameters for TaAs and TaP we performed a density-functional (DFT) calculation for these materials using a generalized gradient approximation \cite{PhysRevLett.77.3865} as implemented in the FPLO code version 48.00-52 \cite{PhysRevB.59.1743}. Using $21^3$ k-points in a box of length $7\cdot 10^{-3}\,\text{\AA}^{-1}$ around the Weyl point, we fitted the DFT data to the generalized model Hamiltonian
\begin{align}
H_G = V\vec{k}\cdot \boldsymbol{\sigma} + \br{\mu + \vec{u}\cdot \vec{k}} \sigma_0
\end{align}
where $\mu$ is the chemical potential and $V=V^T$ a symmetric velocity matrix with chirality $\chi=\text{sign}(\det(V))$. With an accuracy of $10\,meV$, this results for TaAs in $\mu = -8 \,meV$ and (in units of $10^5\, \frac{m}{s}$)
\begin{align}
\vec{u} = \begin{pmatrix}
-0.86\\0.86\\1.44
\end{pmatrix} ,~~~~ V = \begin{pmatrix}
3.38 & 0.84 & 0.88 \\
0.84 & 2.53 & 0.98 \\
0.88 & 0.98 & 2.95
\end{pmatrix}.
\end{align}
and the Weyl node is located at
\begin{align}
	\vec{k}_{W} = \begin{pmatrix} 0.039 \\ 0.511 \\ 0.309 \end{pmatrix}\,\text{\AA}^{-1}.
\end{align}
For TaP we find $\mu = 6 \,meV$ and:
\begin{align}
\vec{k}_W=\begin{pmatrix}
0.032\\0.515\\0.314
\end{pmatrix}, ~\vec{u} = \begin{pmatrix}
-0.86\\0.61\\1.50
\end{pmatrix} ,~V = \begin{pmatrix}
3.08 & 0.69 & 0.74\\
0.69 & 2.29 & 1.11\\
0.74 & 1.11 & 2.82
\end{pmatrix}
\end{align}
The surface state can be solved in the same manner as in the first section. The surface condition changes to $\vec{n}\cdot V\vec{s}=-u_z$ and we get
\begin{align}
\lambda = \kp \cdot \frac{V\br{\vec{s}\times V\vec{n}} - iV\br{V\vec{n}+u_z\vec{s}}}{\left| V\vec{n} \right|^2 - u_z^2}.
\end{align}
A straight-forward calculation yields the BC 
\begin{align}
\vec{\Omega} &= \frac{\text{Im}(\nabla \lambda)\times\text{Re}(\nabla \lambda)}{2\text{Re}^2(\lambda)}\\
&= \frac{ \br{\left| V\vec{n} \right|^2 - u_z^2} \br{V^{-1}\vec{s}\cdot \vec{n}}}{2 \det(V) \br{\kp \cdot \br{V^{-1}\vec{s} \times \vec{n}}}^2} \vec{n}
\end{align}
and finally the BCD
\begin{align}
\vec{D} = \frac{\br{\left| V\vec{n} \right|^2 - u_z^2} \br{V^{-1}\vec{s}\cdot \vec{n}}}{8\pi^2\det(V)k_c \left| V^{-1}\vec{s}\times \vec{n}\right|}\cdot \frac{V\vec{s}+\vec{u}}{1+V^{-1}\vec{s}\cdot \vec{u}}.
\end{align}

\bibliography{InfiniteBC}

\begin{thebibliography}{54}%
\makeatletter
\providecommand \@ifxundefined [1]{%
 \@ifx{#1\undefined}
}%
\providecommand \@ifnum [1]{%
 \ifnum #1\expandafter \@firstoftwo
 \else \expandafter \@secondoftwo
 \fi
}%
\providecommand \@ifx [1]{%
 \ifx #1\expandafter \@firstoftwo
 \else \expandafter \@secondoftwo
 \fi
}%
\providecommand \natexlab [1]{#1}%
\providecommand \enquote  [1]{``#1''}%
\providecommand \bibnamefont  [1]{#1}%
\providecommand \bibfnamefont [1]{#1}%
\providecommand \citenamefont [1]{#1}%
\providecommand \href@noop [0]{\@secondoftwo}%
\providecommand \href [0]{\begingroup \@sanitize@url \@href}%
\providecommand \@href[1]{\@@startlink{#1}\@@href}%
\providecommand \@@href[1]{\endgroup#1\@@endlink}%
\providecommand \@sanitize@url [0]{\catcode `\\12\catcode `\$12\catcode
  `\&12\catcode `\#12\catcode `\^12\catcode `\_12\catcode `\%12\relax}%
\providecommand \@@startlink[1]{}%
\providecommand \@@endlink[0]{}%
\providecommand \url  [0]{\begingroup\@sanitize@url \@url }%
\providecommand \@url [1]{\endgroup\@href {#1}{\urlprefix }}%
\providecommand \urlprefix  [0]{URL }%
\providecommand \Eprint [0]{\href }%
\providecommand \doibase [0]{http://dx.doi.org/}%
\providecommand \selectlanguage [0]{\@gobble}%
\providecommand \bibinfo  [0]{\@secondoftwo}%
\providecommand \bibfield  [0]{\@secondoftwo}%
\providecommand \translation [1]{[#1]}%
\providecommand \BibitemOpen [0]{}%
\providecommand \bibitemStop [0]{}%
\providecommand \bibitemNoStop [0]{.\EOS\space}%
\providecommand \EOS [0]{\spacefactor3000\relax}%
\providecommand \BibitemShut  [1]{\csname bibitem#1\endcsname}%
\let\auto@bib@innerbib\@empty
\bibitem [{\citenamefont {Xiao}\ \emph {et~al.}(2010)\citenamefont {Xiao},
  \citenamefont {Chang},\ and\ \citenamefont {Niu}}]{RevModPhys.82.1959}%
  \BibitemOpen
  \bibfield  {author} {\bibinfo {author} {\bibfnamefont {Di}~\bibnamefont
  {Xiao}}, \bibinfo {author} {\bibfnamefont {Ming-Che}\ \bibnamefont {Chang}},
  \ and\ \bibinfo {author} {\bibfnamefont {Qian}\ \bibnamefont {Niu}},\
  }\bibfield  {title} {\enquote {\bibinfo {title} {Berry phase effects on
  electronic properties},}\ }\href {\doibase 10.1103/RevModPhys.82.1959}
  {\bibfield  {journal} {\bibinfo  {journal} {Rev. Mod. Phys.}\ }\textbf
  {\bibinfo {volume} {82}},\ \bibinfo {pages} {1959--2007} (\bibinfo {year}
  {2010})}\BibitemShut {NoStop}%
\bibitem [{\citenamefont {Armitage}\ \emph {et~al.}(2018)\citenamefont
  {Armitage}, \citenamefont {Mele},\ and\ \citenamefont
  {Vishwanath}}]{RevModPhys.90.015001}%
  \BibitemOpen
  \bibfield  {author} {\bibinfo {author} {\bibfnamefont {N.~P.}\ \bibnamefont
  {Armitage}}, \bibinfo {author} {\bibfnamefont {E.~J.}\ \bibnamefont {Mele}},
  \ and\ \bibinfo {author} {\bibfnamefont {Ashvin}\ \bibnamefont
  {Vishwanath}},\ }\bibfield  {title} {\enquote {\bibinfo {title} {Weyl and
  dirac semimetals in three-dimensional solids},}\ }\href {\doibase
  10.1103/RevModPhys.90.015001} {\bibfield  {journal} {\bibinfo  {journal}
  {Rev. Mod. Phys.}\ }\textbf {\bibinfo {volume} {90}},\ \bibinfo {pages}
  {015001} (\bibinfo {year} {2018})}\BibitemShut {NoStop}%
\bibitem [{\citenamefont {Yan}\ and\ \citenamefont
  {Felser}(2017)}]{annurev-conmatphys-031016-025458}%
  \BibitemOpen
  \bibfield  {author} {\bibinfo {author} {\bibfnamefont {Binghai}\ \bibnamefont
  {Yan}}\ and\ \bibinfo {author} {\bibfnamefont {Claudia}\ \bibnamefont
  {Felser}},\ }\bibfield  {title} {\enquote {\bibinfo {title} {Topological
  materials: Weyl semimetals},}\ }\href {\doibase
  10.1146/annurev-conmatphys-031016-025458} {\bibfield  {journal} {\bibinfo
  {journal} {Annual Review of Condensed Matter Physics}\ }\textbf {\bibinfo
  {volume} {8}},\ \bibinfo {pages} {337--354} (\bibinfo {year} {2017})},\
  \Eprint
  {http://arxiv.org/abs/https://doi.org/10.1146/annurev-conmatphys-031016-025458}
  {https://doi.org/10.1146/annurev-conmatphys-031016-025458} \BibitemShut
  {NoStop}%
\bibitem [{\citenamefont {Deyo}\ \emph {et~al.}(2009)\citenamefont {Deyo},
  \citenamefont {Golub}, \citenamefont {Ivchenko},\ and\ \citenamefont
  {Spivak}}]{deyo2009semiclassical}%
  \BibitemOpen
  \bibfield  {author} {\bibinfo {author} {\bibfnamefont {E}~\bibnamefont
  {Deyo}}, \bibinfo {author} {\bibfnamefont {LE}~\bibnamefont {Golub}},
  \bibinfo {author} {\bibfnamefont {EL}~\bibnamefont {Ivchenko}}, \ and\
  \bibinfo {author} {\bibfnamefont {B}~\bibnamefont {Spivak}},\ }\bibfield
  {title} {\enquote {\bibinfo {title} {Semiclassical theory of the
  photogalvanic effect in non-centrosymmetric systems},}\ }\href@noop {}
  {\bibfield  {journal} {\bibinfo  {journal} {arXiv preprint arXiv:0904.1917}\
  } (\bibinfo {year} {2009})}\BibitemShut {NoStop}%
\bibitem [{\citenamefont {Moore}\ and\ \citenamefont
  {Orenstein}(2010)}]{PhysRevLett.105.026805}%
  \BibitemOpen
  \bibfield  {author} {\bibinfo {author} {\bibfnamefont {J.~E.}\ \bibnamefont
  {Moore}}\ and\ \bibinfo {author} {\bibfnamefont {J.}~\bibnamefont
  {Orenstein}},\ }\bibfield  {title} {\enquote {\bibinfo {title}
  {Confinement-induced berry phase and helicity-dependent photocurrents},}\
  }\href {\doibase 10.1103/PhysRevLett.105.026805} {\bibfield  {journal}
  {\bibinfo  {journal} {Phys. Rev. Lett.}\ }\textbf {\bibinfo {volume} {105}},\
  \bibinfo {pages} {026805} (\bibinfo {year} {2010})}\BibitemShut {NoStop}%
\bibitem [{\citenamefont {Sodemann}\ and\ \citenamefont
  {Fu}(2015)}]{PhysRevLett.115.216806}%
  \BibitemOpen
  \bibfield  {author} {\bibinfo {author} {\bibfnamefont {Inti}\ \bibnamefont
  {Sodemann}}\ and\ \bibinfo {author} {\bibfnamefont {Liang}\ \bibnamefont
  {Fu}},\ }\bibfield  {title} {\enquote {\bibinfo {title} {Quantum nonlinear
  hall effect induced by berry curvature dipole in time-reversal invariant
  materials},}\ }\href {\doibase 10.1103/PhysRevLett.115.216806} {\bibfield
  {journal} {\bibinfo  {journal} {Phys. Rev. Lett.}\ }\textbf {\bibinfo
  {volume} {115}},\ \bibinfo {pages} {216806} (\bibinfo {year}
  {2015})}\BibitemShut {NoStop}%
\bibitem [{\citenamefont {Low}\ \emph {et~al.}(2015)\citenamefont {Low},
  \citenamefont {Jiang},\ and\ \citenamefont {Guinea}}]{PhysRevB.92.235447}%
  \BibitemOpen
  \bibfield  {author} {\bibinfo {author} {\bibfnamefont {Tony}\ \bibnamefont
  {Low}}, \bibinfo {author} {\bibfnamefont {Yongjin}\ \bibnamefont {Jiang}}, \
  and\ \bibinfo {author} {\bibfnamefont {Francisco}\ \bibnamefont {Guinea}},\
  }\bibfield  {title} {\enquote {\bibinfo {title} {Topological currents in
  black phosphorus with broken inversion symmetry},}\ }\href {\doibase
  10.1103/PhysRevB.92.235447} {\bibfield  {journal} {\bibinfo  {journal} {Phys.
  Rev. B}\ }\textbf {\bibinfo {volume} {92}},\ \bibinfo {pages} {235447}
  (\bibinfo {year} {2015})}\BibitemShut {NoStop}%
\bibitem [{\citenamefont {Battilomo}\ \emph {et~al.}(2019)\citenamefont
  {Battilomo}, \citenamefont {Scopigno},\ and\ \citenamefont
  {Ortix}}]{PhysRevLett.123.196403}%
  \BibitemOpen
  \bibfield  {author} {\bibinfo {author} {\bibfnamefont {Raffaele}\
  \bibnamefont {Battilomo}}, \bibinfo {author} {\bibfnamefont {Niccol\'o}\
  \bibnamefont {Scopigno}}, \ and\ \bibinfo {author} {\bibfnamefont {Carmine}\
  \bibnamefont {Ortix}},\ }\bibfield  {title} {\enquote {\bibinfo {title}
  {Berry curvature dipole in strained graphene: A fermi surface warping
  effect},}\ }\href {\doibase 10.1103/PhysRevLett.123.196403} {\bibfield
  {journal} {\bibinfo  {journal} {Phys. Rev. Lett.}\ }\textbf {\bibinfo
  {volume} {123}},\ \bibinfo {pages} {196403} (\bibinfo {year}
  {2019})}\BibitemShut {NoStop}%
\bibitem [{\citenamefont {Xu}\ \emph {et~al.}(2018)\citenamefont {Xu},
  \citenamefont {Ma}, \citenamefont {Shen}, \citenamefont {Fatemi},
  \citenamefont {Wu}, \citenamefont {Chang}, \citenamefont {Chang},
  \citenamefont {Valdivia}, \citenamefont {kit Chan}, \citenamefont {Gibson},
  \citenamefont {Zhou}, \citenamefont {Liu}, \citenamefont {Watanabe},
  \citenamefont {Taniguchi}, \citenamefont {Lin}, \citenamefont {Cava},
  \citenamefont {Fu}, \citenamefont {Gedik},\ and\ \citenamefont
  {Jarillo-Herrero}}]{Xu2018ElectricallySB}%
  \BibitemOpen
  \bibfield  {author} {\bibinfo {author} {\bibfnamefont {Suyang}\ \bibnamefont
  {Xu}}, \bibinfo {author} {\bibfnamefont {Qiong}\ \bibnamefont {Ma}}, \bibinfo
  {author} {\bibfnamefont {Huitao}\ \bibnamefont {Shen}}, \bibinfo {author}
  {\bibfnamefont {Valla}\ \bibnamefont {Fatemi}}, \bibinfo {author}
  {\bibfnamefont {Sanfeng}\ \bibnamefont {Wu}}, \bibinfo {author}
  {\bibfnamefont {Tay-Rong}\ \bibnamefont {Chang}}, \bibinfo {author}
  {\bibfnamefont {Guoqing}\ \bibnamefont {Chang}}, \bibinfo {author}
  {\bibfnamefont {Andr'es M.~Mier}\ \bibnamefont {Valdivia}}, \bibinfo {author}
  {\bibfnamefont {Ching}\ \bibnamefont {kit Chan}}, \bibinfo {author}
  {\bibfnamefont {Quinn~D.}\ \bibnamefont {Gibson}}, \bibinfo {author}
  {\bibfnamefont {Jiadong}\ \bibnamefont {Zhou}}, \bibinfo {author}
  {\bibfnamefont {Zheng}\ \bibnamefont {Liu}}, \bibinfo {author} {\bibfnamefont
  {Kenji}\ \bibnamefont {Watanabe}}, \bibinfo {author} {\bibfnamefont
  {Takashi}\ \bibnamefont {Taniguchi}}, \bibinfo {author} {\bibfnamefont
  {Hsin}\ \bibnamefont {Lin}}, \bibinfo {author} {\bibfnamefont
  {Robert~Joseph}\ \bibnamefont {Cava}}, \bibinfo {author} {\bibfnamefont
  {Liang}\ \bibnamefont {Fu}}, \bibinfo {author} {\bibfnamefont {Nuh}\
  \bibnamefont {Gedik}}, \ and\ \bibinfo {author} {\bibfnamefont {Pablo}\
  \bibnamefont {Jarillo-Herrero}},\ }\bibfield  {title} {\enquote {\bibinfo
  {title} {Electrically switchable berry curvature dipole in the monolayer
  topological insulator wte2},}\ }\href@noop {} {\bibfield  {journal} {\bibinfo
   {journal} {Nature Physics}\ }\textbf {\bibinfo {volume} {14}},\ \bibinfo
  {pages} {900--906} (\bibinfo {year} {2018})}\BibitemShut {NoStop}%
\bibitem [{\citenamefont {You}\ \emph {et~al.}(2018)\citenamefont {You},
  \citenamefont {Fang}, \citenamefont {Xu}, \citenamefont {Kaxiras},\ and\
  \citenamefont {Low}}]{PhysRevB.98.121109}%
  \BibitemOpen
  \bibfield  {author} {\bibinfo {author} {\bibfnamefont {Jhih-Shih}\
  \bibnamefont {You}}, \bibinfo {author} {\bibfnamefont {Shiang}\ \bibnamefont
  {Fang}}, \bibinfo {author} {\bibfnamefont {Su-Yang}\ \bibnamefont {Xu}},
  \bibinfo {author} {\bibfnamefont {Efthimios}\ \bibnamefont {Kaxiras}}, \ and\
  \bibinfo {author} {\bibfnamefont {Tony}\ \bibnamefont {Low}},\ }\bibfield
  {title} {\enquote {\bibinfo {title} {Berry curvature dipole current in the
  transition metal dichalcogenides family},}\ }\href {\doibase
  10.1103/PhysRevB.98.121109} {\bibfield  {journal} {\bibinfo  {journal} {Phys.
  Rev. B}\ }\textbf {\bibinfo {volume} {98}},\ \bibinfo {pages} {121109(R)}
  (\bibinfo {year} {2018})}\BibitemShut {NoStop}%
\bibitem [{\citenamefont {Ma}\ \emph {et~al.}(2019)\citenamefont {Ma},
  \citenamefont {Xu}, \citenamefont {Shen}, \citenamefont {MacNeill},
  \citenamefont {Fatemi}, \citenamefont {Chang}, \citenamefont {Valdivia},
  \citenamefont {Wu}, \citenamefont {Du}, \citenamefont {Hsu} \emph
  {et~al.}}]{ma2019observation}%
  \BibitemOpen
  \bibfield  {author} {\bibinfo {author} {\bibfnamefont {Qiong}\ \bibnamefont
  {Ma}}, \bibinfo {author} {\bibfnamefont {Su-Yang}\ \bibnamefont {Xu}},
  \bibinfo {author} {\bibfnamefont {Huitao}\ \bibnamefont {Shen}}, \bibinfo
  {author} {\bibfnamefont {David}\ \bibnamefont {MacNeill}}, \bibinfo {author}
  {\bibfnamefont {Valla}\ \bibnamefont {Fatemi}}, \bibinfo {author}
  {\bibfnamefont {Tay-Rong}\ \bibnamefont {Chang}}, \bibinfo {author}
  {\bibfnamefont {Andr{\'e}s M~Mier}\ \bibnamefont {Valdivia}}, \bibinfo
  {author} {\bibfnamefont {Sanfeng}\ \bibnamefont {Wu}}, \bibinfo {author}
  {\bibfnamefont {Zongzheng}\ \bibnamefont {Du}}, \bibinfo {author}
  {\bibfnamefont {Chuang-Han}\ \bibnamefont {Hsu}},  \emph {et~al.},\
  }\bibfield  {title} {\enquote {\bibinfo {title} {Observation of the nonlinear
  hall effect under time-reversal-symmetric conditions},}\ }\href@noop {}
  {\bibfield  {journal} {\bibinfo  {journal} {Nature}\ }\textbf {\bibinfo
  {volume} {565}},\ \bibinfo {pages} {337--342} (\bibinfo {year}
  {2019})}\BibitemShut {NoStop}%
\bibitem [{\citenamefont {Kang}\ \emph {et~al.}(2019)\citenamefont {Kang},
  \citenamefont {Li}, \citenamefont {Sohn}, \citenamefont {Shan},\ and\
  \citenamefont {Mak}}]{kang2019nonlinear}%
  \BibitemOpen
  \bibfield  {author} {\bibinfo {author} {\bibfnamefont {Kaifei}\ \bibnamefont
  {Kang}}, \bibinfo {author} {\bibfnamefont {Tingxin}\ \bibnamefont {Li}},
  \bibinfo {author} {\bibfnamefont {Egon}\ \bibnamefont {Sohn}}, \bibinfo
  {author} {\bibfnamefont {Jie}\ \bibnamefont {Shan}}, \ and\ \bibinfo {author}
  {\bibfnamefont {Kin~Fai}\ \bibnamefont {Mak}},\ }\bibfield  {title} {\enquote
  {\bibinfo {title} {Nonlinear anomalous hall effect in few-layer wte 2},}\
  }\href@noop {} {\bibfield  {journal} {\bibinfo  {journal} {Nature materials}\
  }\textbf {\bibinfo {volume} {18}},\ \bibinfo {pages} {324--328} (\bibinfo
  {year} {2019})}\BibitemShut {NoStop}%
\bibitem [{\citenamefont {Facio}\ \emph {et~al.}(2018)\citenamefont {Facio},
  \citenamefont {Efremov}, \citenamefont {Koepernik}, \citenamefont {You},
  \citenamefont {Sodemann},\ and\ \citenamefont {van~den
  Brink}}]{PhysRevLett.121.246403}%
  \BibitemOpen
  \bibfield  {author} {\bibinfo {author} {\bibfnamefont {Jorge~I.}\
  \bibnamefont {Facio}}, \bibinfo {author} {\bibfnamefont {Dmitri}\
  \bibnamefont {Efremov}}, \bibinfo {author} {\bibfnamefont {Klaus}\
  \bibnamefont {Koepernik}}, \bibinfo {author} {\bibfnamefont {Jhih-Shih}\
  \bibnamefont {You}}, \bibinfo {author} {\bibfnamefont {Inti}\ \bibnamefont
  {Sodemann}}, \ and\ \bibinfo {author} {\bibfnamefont {Jeroen}\ \bibnamefont
  {van~den Brink}},\ }\bibfield  {title} {\enquote {\bibinfo {title} {Strongly
  enhanced berry dipole at topological phase transitions in bitei},}\ }\href
  {\doibase 10.1103/PhysRevLett.121.246403} {\bibfield  {journal} {\bibinfo
  {journal} {Phys. Rev. Lett.}\ }\textbf {\bibinfo {volume} {121}},\ \bibinfo
  {pages} {246403} (\bibinfo {year} {2018})}\BibitemShut {NoStop}%
\bibitem [{\citenamefont {Zhang}\ \emph {et~al.}(2018)\citenamefont {Zhang},
  \citenamefont {Sun},\ and\ \citenamefont {Yan}}]{PhysRevB.97.041101}%
  \BibitemOpen
  \bibfield  {author} {\bibinfo {author} {\bibfnamefont {Yang}\ \bibnamefont
  {Zhang}}, \bibinfo {author} {\bibfnamefont {Yan}\ \bibnamefont {Sun}}, \ and\
  \bibinfo {author} {\bibfnamefont {Binghai}\ \bibnamefont {Yan}},\ }\bibfield
  {title} {\enquote {\bibinfo {title} {Berry curvature dipole in weyl semimetal
  materials: An ab initio study},}\ }\href {\doibase
  10.1103/PhysRevB.97.041101} {\bibfield  {journal} {\bibinfo  {journal} {Phys.
  Rev. B}\ }\textbf {\bibinfo {volume} {97}},\ \bibinfo {pages} {041101(R)}
  (\bibinfo {year} {2018})}\BibitemShut {NoStop}%
\bibitem [{\citenamefont {Shvetsov}\ \emph {et~al.}(2019)\citenamefont
  {Shvetsov}, \citenamefont {Esin}, \citenamefont {Timonina}, \citenamefont
  {Kolesnikov},\ and\ \citenamefont {Deviatov}}]{shvetsov2019nonlinear}%
  \BibitemOpen
  \bibfield  {author} {\bibinfo {author} {\bibfnamefont {Oleg~Olegovich}\
  \bibnamefont {Shvetsov}}, \bibinfo {author} {\bibfnamefont
  {Varnava~Denisovich}\ \bibnamefont {Esin}}, \bibinfo {author} {\bibfnamefont
  {Anna~Vladimirovna}\ \bibnamefont {Timonina}}, \bibinfo {author}
  {\bibfnamefont {Nikolay~Nikolaevich}\ \bibnamefont {Kolesnikov}}, \ and\
  \bibinfo {author} {\bibfnamefont {EV}~\bibnamefont {Deviatov}},\ }\bibfield
  {title} {\enquote {\bibinfo {title} {Nonlinear hall effect in
  three-dimensional weyl and dirac semimetals},}\ }\href@noop {} {\bibfield
  {journal} {\bibinfo  {journal} {JETP Letters}\ }\textbf {\bibinfo {volume}
  {109}},\ \bibinfo {pages} {715--721} (\bibinfo {year} {2019})}\BibitemShut
  {NoStop}%
\bibitem [{\citenamefont {Matsyshyn}\ and\ \citenamefont
  {Sodemann}(2019)}]{PhysRevLett.123.246602}%
  \BibitemOpen
  \bibfield  {author} {\bibinfo {author} {\bibfnamefont {O.}~\bibnamefont
  {Matsyshyn}}\ and\ \bibinfo {author} {\bibfnamefont {I.}~\bibnamefont
  {Sodemann}},\ }\bibfield  {title} {\enquote {\bibinfo {title} {Nonlinear hall
  acceleration and the quantum rectification sum rule},}\ }\href {\doibase
  10.1103/PhysRevLett.123.246602} {\bibfield  {journal} {\bibinfo  {journal}
  {Phys. Rev. Lett.}\ }\textbf {\bibinfo {volume} {123}},\ \bibinfo {pages}
  {246602} (\bibinfo {year} {2019})}\BibitemShut {NoStop}%
\bibitem [{\citenamefont {Nielsen}\ and\ \citenamefont
  {Ninomiya}(1981)}]{NIELSEN1981219}%
  \BibitemOpen
  \bibfield  {author} {\bibinfo {author} {\bibfnamefont {H.B.}\ \bibnamefont
  {Nielsen}}\ and\ \bibinfo {author} {\bibfnamefont {M.}~\bibnamefont
  {Ninomiya}},\ }\bibfield  {title} {\enquote {\bibinfo {title} {A no-go
  theorem for regularizing chiral fermions},}\ }\href {\doibase
  https://doi.org/10.1016/0370-2693(81)91026-1} {\bibfield  {journal} {\bibinfo
   {journal} {Physics Letters B}\ }\textbf {\bibinfo {volume} {105}},\ \bibinfo
  {pages} {219 -- 223} (\bibinfo {year} {1981})}\BibitemShut {NoStop}%
\bibitem [{\citenamefont {Li}\ and\ \citenamefont
  {Andreev}(2015)}]{PhysRevB.92.201107}%
  \BibitemOpen
  \bibfield  {author} {\bibinfo {author} {\bibfnamefont {Songci}\ \bibnamefont
  {Li}}\ and\ \bibinfo {author} {\bibfnamefont {A.~V.}\ \bibnamefont
  {Andreev}},\ }\bibfield  {title} {\enquote {\bibinfo {title} {Spiraling fermi
  arcs in weyl materials},}\ }\href {\doibase 10.1103/PhysRevB.92.201107}
  {\bibfield  {journal} {\bibinfo  {journal} {Phys. Rev. B}\ }\textbf {\bibinfo
  {volume} {92}},\ \bibinfo {pages} {201107(R)} (\bibinfo {year}
  {2015})}\BibitemShut {NoStop}%
\bibitem [{\citenamefont {Witten}(2016)}]{witten2016three}%
  \BibitemOpen
  \bibfield  {author} {\bibinfo {author} {\bibfnamefont {Edward}\ \bibnamefont
  {Witten}},\ }\bibfield  {title} {\enquote {\bibinfo {title} {Three lectures
  on topological phases of matter},}\ }\href@noop {} {\bibfield  {journal}
  {\bibinfo  {journal} {La Rivista del Nuovo Cimento}\ }\textbf {\bibinfo
  {volume} {39}},\ \bibinfo {pages} {313--370} (\bibinfo {year}
  {2016})}\BibitemShut {NoStop}%
\bibitem [{\citenamefont {Hashimoto}\ \emph {et~al.}(2017)\citenamefont
  {Hashimoto}, \citenamefont {Kimura},\ and\ \citenamefont
  {Wu}}]{10.1093/ptep/ptx053}%
  \BibitemOpen
  \bibfield  {author} {\bibinfo {author} {\bibfnamefont {Koji}\ \bibnamefont
  {Hashimoto}}, \bibinfo {author} {\bibfnamefont {Taro}\ \bibnamefont
  {Kimura}}, \ and\ \bibinfo {author} {\bibfnamefont {Xi}~\bibnamefont {Wu}},\
  }\bibfield  {title} {\enquote {\bibinfo {title} {{Boundary conditions of Weyl
  semimetals}},}\ }\href {\doibase 10.1093/ptep/ptx053} {\bibfield  {journal}
  {\bibinfo  {journal} {Progress of Theoretical and Experimental Physics}\
  }\textbf {\bibinfo {volume} {2017}} (\bibinfo {year} {2017}),\
  10.1093/ptep/ptx053},\ \bibinfo {note} {053I01}\BibitemShut {NoStop}%
\bibitem [{\citenamefont {Seradjeh}\ and\ \citenamefont
  {Vennettilli}(2018)}]{PhysRevB.97.075132}%
  \BibitemOpen
  \bibfield  {author} {\bibinfo {author} {\bibfnamefont {Babak}\ \bibnamefont
  {Seradjeh}}\ and\ \bibinfo {author} {\bibfnamefont {Michael}\ \bibnamefont
  {Vennettilli}},\ }\bibfield  {title} {\enquote {\bibinfo {title} {Surface
  spectra of weyl semimetals through self-adjoint extensions},}\ }\href
  {\doibase 10.1103/PhysRevB.97.075132} {\bibfield  {journal} {\bibinfo
  {journal} {Phys. Rev. B}\ }\textbf {\bibinfo {volume} {97}},\ \bibinfo
  {pages} {075132} (\bibinfo {year} {2018})}\BibitemShut {NoStop}%
\bibitem [{\citenamefont {Burrello}\ \emph {et~al.}(2019)\citenamefont
  {Burrello}, \citenamefont {Guadagnini}, \citenamefont {Lepori},\ and\
  \citenamefont {Mintchev}}]{PhysRevB.100.155131}%
  \BibitemOpen
  \bibfield  {author} {\bibinfo {author} {\bibfnamefont {Michele}\ \bibnamefont
  {Burrello}}, \bibinfo {author} {\bibfnamefont {Enore}\ \bibnamefont
  {Guadagnini}}, \bibinfo {author} {\bibfnamefont {Luca}\ \bibnamefont
  {Lepori}}, \ and\ \bibinfo {author} {\bibfnamefont {Mihail}\ \bibnamefont
  {Mintchev}},\ }\bibfield  {title} {\enquote {\bibinfo {title} {Field theory
  approach to the quantum transport in weyl semimetals},}\ }\href {\doibase
  10.1103/PhysRevB.100.155131} {\bibfield  {journal} {\bibinfo  {journal}
  {Phys. Rev. B}\ }\textbf {\bibinfo {volume} {100}},\ \bibinfo {pages}
  {155131} (\bibinfo {year} {2019})}\BibitemShut {NoStop}%
\bibitem [{Sup()}]{Supp}%
  \BibitemOpen
  \href@noop {} {\bibinfo  {journal} {See Supplemental Material for the full
  surface wave functions for semi-infinite and finite systems, an explicit
  interface between vacuum and Weyl semimetal, and the parameters for TaAs and
  TaP}\ }\BibitemShut {NoStop}%
\bibitem [{\citenamefont {Soluyanov}\ \emph {et~al.}(2015)\citenamefont
  {Soluyanov}, \citenamefont {Gresch}, \citenamefont {Wang}, \citenamefont
  {Wu}, \citenamefont {Troyer}, \citenamefont {Dai},\ and\ \citenamefont
  {Bernevig}}]{soluyanov2015type}%
  \BibitemOpen
\bibfield  {journal} {  }\bibfield  {author} {\bibinfo {author} {\bibfnamefont
  {Alexey~A}\ \bibnamefont {Soluyanov}}, \bibinfo {author} {\bibfnamefont
  {Dominik}\ \bibnamefont {Gresch}}, \bibinfo {author} {\bibfnamefont {Zhijun}\
  \bibnamefont {Wang}}, \bibinfo {author} {\bibfnamefont {QuanSheng}\
  \bibnamefont {Wu}}, \bibinfo {author} {\bibfnamefont {Matthias}\ \bibnamefont
  {Troyer}}, \bibinfo {author} {\bibfnamefont {Xi}~\bibnamefont {Dai}}, \ and\
  \bibinfo {author} {\bibfnamefont {B~Andrei}\ \bibnamefont {Bernevig}},\
  }\bibfield  {title} {\enquote {\bibinfo {title} {Type-ii weyl semimetals},}\
  }\href@noop {} {\bibfield  {journal} {\bibinfo  {journal} {Nature}\ }\textbf
  {\bibinfo {volume} {527}},\ \bibinfo {pages} {495--498} (\bibinfo {year}
  {2015})}\BibitemShut {NoStop}%
\bibitem [{\citenamefont {Perdew}\ \emph {et~al.}(1996)\citenamefont {Perdew},
  \citenamefont {Burke},\ and\ \citenamefont
  {Ernzerhof}}]{PhysRevLett.77.3865}%
  \BibitemOpen
  \bibfield  {author} {\bibinfo {author} {\bibfnamefont {John~P.}\ \bibnamefont
  {Perdew}}, \bibinfo {author} {\bibfnamefont {Kieron}\ \bibnamefont {Burke}},
  \ and\ \bibinfo {author} {\bibfnamefont {Matthias}\ \bibnamefont
  {Ernzerhof}},\ }\bibfield  {title} {\enquote {\bibinfo {title} {Generalized
  gradient approximation made simple},}\ }\href {\doibase
  10.1103/PhysRevLett.77.3865} {\bibfield  {journal} {\bibinfo  {journal}
  {Phys. Rev. Lett.}\ }\textbf {\bibinfo {volume} {77}},\ \bibinfo {pages}
  {3865--3868} (\bibinfo {year} {1996})}\BibitemShut {NoStop}%
\bibitem [{\citenamefont {Koepernik}\ and\ \citenamefont
  {Eschrig}(1999)}]{PhysRevB.59.1743}%
  \BibitemOpen
  \bibfield  {author} {\bibinfo {author} {\bibfnamefont {Klaus}\ \bibnamefont
  {Koepernik}}\ and\ \bibinfo {author} {\bibfnamefont {Helmut}\ \bibnamefont
  {Eschrig}},\ }\bibfield  {title} {\enquote {\bibinfo {title} {Full-potential
  nonorthogonal local-orbital minimum-basis band-structure scheme},}\ }\href
  {\doibase 10.1103/PhysRevB.59.1743} {\bibfield  {journal} {\bibinfo
  {journal} {Phys. Rev. B}\ }\textbf {\bibinfo {volume} {59}},\ \bibinfo
  {pages} {1743--1757} (\bibinfo {year} {1999})}\BibitemShut {NoStop}%
\bibitem [{\citenamefont {Weng}\ \emph {et~al.}(2015)\citenamefont {Weng},
  \citenamefont {Fang}, \citenamefont {Fang}, \citenamefont {Bernevig},\ and\
  \citenamefont {Dai}}]{PhysRevX.5.011029}%
  \BibitemOpen
  \bibfield  {author} {\bibinfo {author} {\bibfnamefont {Hongming}\
  \bibnamefont {Weng}}, \bibinfo {author} {\bibfnamefont {Chen}\ \bibnamefont
  {Fang}}, \bibinfo {author} {\bibfnamefont {Zhong}\ \bibnamefont {Fang}},
  \bibinfo {author} {\bibfnamefont {B.~Andrei}\ \bibnamefont {Bernevig}}, \
  and\ \bibinfo {author} {\bibfnamefont {Xi}~\bibnamefont {Dai}},\ }\bibfield
  {title} {\enquote {\bibinfo {title} {Weyl semimetal phase in
  noncentrosymmetric transition-metal monophosphides},}\ }\href {\doibase
  10.1103/PhysRevX.5.011029} {\bibfield  {journal} {\bibinfo  {journal} {Phys.
  Rev. X}\ }\textbf {\bibinfo {volume} {5}},\ \bibinfo {pages} {011029}
  (\bibinfo {year} {2015})}\BibitemShut {NoStop}%
\bibitem [{\citenamefont {Lv}\ \emph {et~al.}(2015)\citenamefont {Lv},
  \citenamefont {Xu}, \citenamefont {Weng}, \citenamefont {Ma}, \citenamefont
  {Richard}, \citenamefont {Huang}, \citenamefont {Zhao}, \citenamefont {Chen},
  \citenamefont {Matt}, \citenamefont {Bisti} \emph
  {et~al.}}]{lv2015observation}%
  \BibitemOpen
  \bibfield  {author} {\bibinfo {author} {\bibfnamefont {BQ}~\bibnamefont
  {Lv}}, \bibinfo {author} {\bibfnamefont {N}~\bibnamefont {Xu}}, \bibinfo
  {author} {\bibfnamefont {HM}~\bibnamefont {Weng}}, \bibinfo {author}
  {\bibfnamefont {JZ}~\bibnamefont {Ma}}, \bibinfo {author} {\bibfnamefont
  {P}~\bibnamefont {Richard}}, \bibinfo {author} {\bibfnamefont
  {XC}~\bibnamefont {Huang}}, \bibinfo {author} {\bibfnamefont
  {LX}~\bibnamefont {Zhao}}, \bibinfo {author} {\bibfnamefont {GF}~\bibnamefont
  {Chen}}, \bibinfo {author} {\bibfnamefont {CE}~\bibnamefont {Matt}}, \bibinfo
  {author} {\bibfnamefont {F}~\bibnamefont {Bisti}},  \emph {et~al.},\
  }\bibfield  {title} {\enquote {\bibinfo {title} {Observation of weyl nodes in
  taas},}\ }\href@noop {} {\bibfield  {journal} {\bibinfo  {journal} {Nature
  Physics}\ }\textbf {\bibinfo {volume} {11}},\ \bibinfo {pages} {724--727}
  (\bibinfo {year} {2015})}\BibitemShut {NoStop}%
\bibitem [{\citenamefont {Son}\ \emph {et~al.}(2019)\citenamefont {Son},
  \citenamefont {Kim}, \citenamefont {Ahn}, \citenamefont {Lee},\ and\
  \citenamefont {Lee}}]{PhysRevLett.123.036806}%
  \BibitemOpen
  \bibfield  {author} {\bibinfo {author} {\bibfnamefont {Joolee}\ \bibnamefont
  {Son}}, \bibinfo {author} {\bibfnamefont {Kyung-Han}\ \bibnamefont {Kim}},
  \bibinfo {author} {\bibfnamefont {Y.~H.}\ \bibnamefont {Ahn}}, \bibinfo
  {author} {\bibfnamefont {Hyun-Woo}\ \bibnamefont {Lee}}, \ and\ \bibinfo
  {author} {\bibfnamefont {Jieun}\ \bibnamefont {Lee}},\ }\bibfield  {title}
  {\enquote {\bibinfo {title} {Strain engineering of the berry curvature dipole
  and valley magnetization in monolayer ${\mathrm{mos}}_{2}$},}\ }\href
  {\doibase 10.1103/PhysRevLett.123.036806} {\bibfield  {journal} {\bibinfo
  {journal} {Phys. Rev. Lett.}\ }\textbf {\bibinfo {volume} {123}},\ \bibinfo
  {pages} {036806} (\bibinfo {year} {2019})}\BibitemShut {NoStop}%
\bibitem [{\citenamefont {Yu}\ \emph {et~al.}(2019)\citenamefont {Yu},
  \citenamefont {Zhu}, \citenamefont {You}, \citenamefont {Low},\ and\
  \citenamefont {Su}}]{PhysRevB.99.201410}%
  \BibitemOpen
  \bibfield  {author} {\bibinfo {author} {\bibfnamefont {Xiao-Qin}\
  \bibnamefont {Yu}}, \bibinfo {author} {\bibfnamefont {Zhen-Gang}\
  \bibnamefont {Zhu}}, \bibinfo {author} {\bibfnamefont {Jhih-Shih}\
  \bibnamefont {You}}, \bibinfo {author} {\bibfnamefont {Tony}\ \bibnamefont
  {Low}}, \ and\ \bibinfo {author} {\bibfnamefont {Gang}\ \bibnamefont {Su}},\
  }\bibfield  {title} {\enquote {\bibinfo {title} {Topological nonlinear
  anomalous nernst effect in strained transition metal dichalcogenides},}\
  }\href {\doibase 10.1103/PhysRevB.99.201410} {\bibfield  {journal} {\bibinfo
  {journal} {Phys. Rev. B}\ }\textbf {\bibinfo {volume} {99}},\ \bibinfo
  {pages} {201410(R)} (\bibinfo {year} {2019})}\BibitemShut {NoStop}%
\bibitem [{\citenamefont {Zeng}\ \emph {et~al.}(2020)\citenamefont {Zeng},
  \citenamefont {Nandy},\ and\ \citenamefont
  {Tewari}}]{PhysRevResearch.2.032066}%
  \BibitemOpen
  \bibfield  {author} {\bibinfo {author} {\bibfnamefont {Chuanchang}\
  \bibnamefont {Zeng}}, \bibinfo {author} {\bibfnamefont {Snehasish}\
  \bibnamefont {Nandy}}, \ and\ \bibinfo {author} {\bibfnamefont {Sumanta}\
  \bibnamefont {Tewari}},\ }\bibfield  {title} {\enquote {\bibinfo {title}
  {Fundamental relations for anomalous thermoelectric transport coefficients in
  the nonlinear regime},}\ }\href {\doibase 10.1103/PhysRevResearch.2.032066}
  {\bibfield  {journal} {\bibinfo  {journal} {Phys. Rev. Research}\ }\textbf
  {\bibinfo {volume} {2}},\ \bibinfo {pages} {032066(R)} (\bibinfo {year}
  {2020})}\BibitemShut {NoStop}%
\bibitem [{\citenamefont {Sundaram}\ and\ \citenamefont
  {Niu}(1999)}]{PhysRevB.59.14915}%
  \BibitemOpen
  \bibfield  {author} {\bibinfo {author} {\bibfnamefont {Ganesh}\ \bibnamefont
  {Sundaram}}\ and\ \bibinfo {author} {\bibfnamefont {Qian}\ \bibnamefont
  {Niu}},\ }\bibfield  {title} {\enquote {\bibinfo {title} {Wave-packet
  dynamics in slowly perturbed crystals: Gradient corrections and berry-phase
  effects},}\ }\href {\doibase 10.1103/PhysRevB.59.14915} {\bibfield  {journal}
  {\bibinfo  {journal} {Phys. Rev. B}\ }\textbf {\bibinfo {volume} {59}},\
  \bibinfo {pages} {14915--14925} (\bibinfo {year} {1999})}\BibitemShut
  {NoStop}%
\bibitem [{\citenamefont {Flicker}\ \emph {et~al.}(2018)\citenamefont
  {Flicker}, \citenamefont {de~Juan}, \citenamefont {Bradlyn}, \citenamefont
  {Morimoto}, \citenamefont {Vergniory},\ and\ \citenamefont
  {Grushin}}]{PhysRevB.98.155145}%
  \BibitemOpen
  \bibfield  {author} {\bibinfo {author} {\bibfnamefont {Felix}\ \bibnamefont
  {Flicker}}, \bibinfo {author} {\bibfnamefont {Fernando}\ \bibnamefont
  {de~Juan}}, \bibinfo {author} {\bibfnamefont {Barry}\ \bibnamefont
  {Bradlyn}}, \bibinfo {author} {\bibfnamefont {Takahiro}\ \bibnamefont
  {Morimoto}}, \bibinfo {author} {\bibfnamefont {Maia~G.}\ \bibnamefont
  {Vergniory}}, \ and\ \bibinfo {author} {\bibfnamefont {Adolfo~G.}\
  \bibnamefont {Grushin}},\ }\bibfield  {title} {\enquote {\bibinfo {title}
  {Chiral optical response of multifold fermions},}\ }\href {\doibase
  10.1103/PhysRevB.98.155145} {\bibfield  {journal} {\bibinfo  {journal} {Phys.
  Rev. B}\ }\textbf {\bibinfo {volume} {98}},\ \bibinfo {pages} {155145}
  (\bibinfo {year} {2018})}\BibitemShut {NoStop}%
\bibitem [{\citenamefont {Knoll}\ \emph {et~al.}(2020)\citenamefont {Knoll},
  \citenamefont {Timm},\ and\ \citenamefont {Meng}}]{PhysRevB.101.201402}%
  \BibitemOpen
  \bibfield  {author} {\bibinfo {author} {\bibfnamefont {Andy}\ \bibnamefont
  {Knoll}}, \bibinfo {author} {\bibfnamefont {Carsten}\ \bibnamefont {Timm}}, \
  and\ \bibinfo {author} {\bibfnamefont {Tobias}\ \bibnamefont {Meng}},\
  }\bibfield  {title} {\enquote {\bibinfo {title} {Negative longitudinal
  magnetoconductance at weak fields in weyl semimetals},}\ }\href {\doibase
  10.1103/PhysRevB.101.201402} {\bibfield  {journal} {\bibinfo  {journal}
  {Phys. Rev. B}\ }\textbf {\bibinfo {volume} {101}},\ \bibinfo {pages}
  {201402(R)} (\bibinfo {year} {2020})}\BibitemShut {NoStop}%
\bibitem [{\citenamefont {Li}\ \emph {et~al.}(2021)\citenamefont {Li},
  \citenamefont {Heinonen}, \citenamefont {Burkov},\ and\ \citenamefont
  {Zhang}}]{PhysRevB.103.045105}%
  \BibitemOpen
  \bibfield  {author} {\bibinfo {author} {\bibfnamefont {Rui-Hao}\ \bibnamefont
  {Li}}, \bibinfo {author} {\bibfnamefont {Olle~G.}\ \bibnamefont {Heinonen}},
  \bibinfo {author} {\bibfnamefont {Anton~A.}\ \bibnamefont {Burkov}}, \ and\
  \bibinfo {author} {\bibfnamefont {Steven S.-L.}\ \bibnamefont {Zhang}},\
  }\bibfield  {title} {\enquote {\bibinfo {title} {Nonlinear hall effect in
  weyl semimetals induced by chiral anomaly},}\ }\href {\doibase
  10.1103/PhysRevB.103.045105} {\bibfield  {journal} {\bibinfo  {journal}
  {Phys. Rev. B}\ }\textbf {\bibinfo {volume} {103}},\ \bibinfo {pages}
  {045105} (\bibinfo {year} {2021})}\BibitemShut {NoStop}%
\bibitem [{\citenamefont {Gorbar}\ \emph {et~al.}(2016)\citenamefont {Gorbar},
  \citenamefont {Miransky}, \citenamefont {Shovkovy},\ and\ \citenamefont
  {Sukhachov}}]{PhysRevB.93.235127}%
  \BibitemOpen
  \bibfield  {author} {\bibinfo {author} {\bibfnamefont {E.~V.}\ \bibnamefont
  {Gorbar}}, \bibinfo {author} {\bibfnamefont {V.~A.}\ \bibnamefont
  {Miransky}}, \bibinfo {author} {\bibfnamefont {I.~A.}\ \bibnamefont
  {Shovkovy}}, \ and\ \bibinfo {author} {\bibfnamefont {P.~O.}\ \bibnamefont
  {Sukhachov}},\ }\bibfield  {title} {\enquote {\bibinfo {title} {Origin of
  dissipative fermi arc transport in weyl semimetals},}\ }\href {\doibase
  10.1103/PhysRevB.93.235127} {\bibfield  {journal} {\bibinfo  {journal} {Phys.
  Rev. B}\ }\textbf {\bibinfo {volume} {93}},\ \bibinfo {pages} {235127}
  (\bibinfo {year} {2016})}\BibitemShut {NoStop}%
\bibitem [{\citenamefont {Slager}\ \emph {et~al.}(2017)\citenamefont {Slager},
  \citenamefont {Juri\ifmmode \check{c}\else \v{c}\fi{}i\ifmmode~\acute{c}\else
  \'{c}\fi{}},\ and\ \citenamefont {Roy}}]{PhysRevB.96.201401}%
  \BibitemOpen
  \bibfield  {author} {\bibinfo {author} {\bibfnamefont {Robert-Jan}\
  \bibnamefont {Slager}}, \bibinfo {author} {\bibfnamefont {Vladimir}\
  \bibnamefont {Juri\ifmmode \check{c}\else \v{c}\fi{}i\ifmmode~\acute{c}\else
  \'{c}\fi{}}}, \ and\ \bibinfo {author} {\bibfnamefont {Bitan}\ \bibnamefont
  {Roy}},\ }\bibfield  {title} {\enquote {\bibinfo {title} {Dissolution of
  topological fermi arcs in a dirty weyl semimetal},}\ }\href {\doibase
  10.1103/PhysRevB.96.201401} {\bibfield  {journal} {\bibinfo  {journal} {Phys.
  Rev. B}\ }\textbf {\bibinfo {volume} {96}},\ \bibinfo {pages} {201401(R)}
  (\bibinfo {year} {2017})}\BibitemShut {NoStop}%
\bibitem [{\citenamefont {Wilson}\ \emph {et~al.}(2018)\citenamefont {Wilson},
  \citenamefont {Pixley}, \citenamefont {Huse}, \citenamefont {Refael},\ and\
  \citenamefont {Das~Sarma}}]{PhysRevB.97.235108}%
  \BibitemOpen
  \bibfield  {author} {\bibinfo {author} {\bibfnamefont {Justin~H.}\
  \bibnamefont {Wilson}}, \bibinfo {author} {\bibfnamefont {J.~H.}\
  \bibnamefont {Pixley}}, \bibinfo {author} {\bibfnamefont {David~A.}\
  \bibnamefont {Huse}}, \bibinfo {author} {\bibfnamefont {Gil}\ \bibnamefont
  {Refael}}, \ and\ \bibinfo {author} {\bibfnamefont {S.}~\bibnamefont
  {Das~Sarma}},\ }\bibfield  {title} {\enquote {\bibinfo {title} {Do the
  surface fermi arcs in weyl semimetals survive disorder?}}\ }\href {\doibase
  10.1103/PhysRevB.97.235108} {\bibfield  {journal} {\bibinfo  {journal} {Phys.
  Rev. B}\ }\textbf {\bibinfo {volume} {97}},\ \bibinfo {pages} {235108}
  (\bibinfo {year} {2018})}\BibitemShut {NoStop}%
\bibitem [{\citenamefont {Roy}\ \emph {et~al.}(2018)\citenamefont {Roy},
  \citenamefont {Slager},\ and\ \citenamefont {Juri\ifmmode \check{c}\else
  \v{c}\fi{}i\ifmmode~\acute{c}\else \'{c}\fi{}}}]{PhysRevX.8.031076}%
  \BibitemOpen
  \bibfield  {author} {\bibinfo {author} {\bibfnamefont {Bitan}\ \bibnamefont
  {Roy}}, \bibinfo {author} {\bibfnamefont {Robert-Jan}\ \bibnamefont
  {Slager}}, \ and\ \bibinfo {author} {\bibfnamefont {Vladimir}\ \bibnamefont
  {Juri\ifmmode \check{c}\else \v{c}\fi{}i\ifmmode~\acute{c}\else
  \'{c}\fi{}}},\ }\bibfield  {title} {\enquote {\bibinfo {title} {Global phase
  diagram of a dirty weyl liquid and emergent superuniversality},}\ }\href
  {\doibase 10.1103/PhysRevX.8.031076} {\bibfield  {journal} {\bibinfo
  {journal} {Phys. Rev. X}\ }\textbf {\bibinfo {volume} {8}},\ \bibinfo {pages}
  {031076} (\bibinfo {year} {2018})}\BibitemShut {NoStop}%
\bibitem [{\citenamefont {K\"onig}\ \emph {et~al.}(2019)\citenamefont
  {K\"onig}, \citenamefont {Dzero}, \citenamefont {Levchenko},\ and\
  \citenamefont {Pesin}}]{PhysRevB.99.155404}%
  \BibitemOpen
  \bibfield  {author} {\bibinfo {author} {\bibfnamefont {E.~J.}\ \bibnamefont
  {K\"onig}}, \bibinfo {author} {\bibfnamefont {M.}~\bibnamefont {Dzero}},
  \bibinfo {author} {\bibfnamefont {A.}~\bibnamefont {Levchenko}}, \ and\
  \bibinfo {author} {\bibfnamefont {D.~A.}\ \bibnamefont {Pesin}},\ }\bibfield
  {title} {\enquote {\bibinfo {title} {Gyrotropic hall effect in berry-curved
  materials},}\ }\href {\doibase 10.1103/PhysRevB.99.155404} {\bibfield
  {journal} {\bibinfo  {journal} {Phys. Rev. B}\ }\textbf {\bibinfo {volume}
  {99}},\ \bibinfo {pages} {155404} (\bibinfo {year} {2019})}\BibitemShut
  {NoStop}%
\bibitem [{\citenamefont {Xiao}\ \emph {et~al.}(2019)\citenamefont {Xiao},
  \citenamefont {Du},\ and\ \citenamefont {Niu}}]{PhysRevB.100.165422}%
  \BibitemOpen
  \bibfield  {author} {\bibinfo {author} {\bibfnamefont {Cong}\ \bibnamefont
  {Xiao}}, \bibinfo {author} {\bibfnamefont {Z.~Z.}\ \bibnamefont {Du}}, \ and\
  \bibinfo {author} {\bibfnamefont {Qian}\ \bibnamefont {Niu}},\ }\bibfield
  {title} {\enquote {\bibinfo {title} {Theory of nonlinear hall effects:
  Modified semiclassics from quantum kinetics},}\ }\href {\doibase
  10.1103/PhysRevB.100.165422} {\bibfield  {journal} {\bibinfo  {journal}
  {Phys. Rev. B}\ }\textbf {\bibinfo {volume} {100}},\ \bibinfo {pages}
  {165422} (\bibinfo {year} {2019})}\BibitemShut {NoStop}%
\bibitem [{\citenamefont {Nandy}\ and\ \citenamefont
  {Sodemann}(2019)}]{PhysRevB.100.195117}%
  \BibitemOpen
  \bibfield  {author} {\bibinfo {author} {\bibfnamefont {S.}~\bibnamefont
  {Nandy}}\ and\ \bibinfo {author} {\bibfnamefont {Inti}\ \bibnamefont
  {Sodemann}},\ }\bibfield  {title} {\enquote {\bibinfo {title} {Symmetry and
  quantum kinetics of the nonlinear hall effect},}\ }\href {\doibase
  10.1103/PhysRevB.100.195117} {\bibfield  {journal} {\bibinfo  {journal}
  {Phys. Rev. B}\ }\textbf {\bibinfo {volume} {100}},\ \bibinfo {pages}
  {195117} (\bibinfo {year} {2019})}\BibitemShut {NoStop}%
\bibitem [{\citenamefont {Du}\ \emph {et~al.}(2019)\citenamefont {Du},
  \citenamefont {Wang}, \citenamefont {Li}, \citenamefont {Lu},\ and\
  \citenamefont {Xie}}]{du2019disorder}%
  \BibitemOpen
  \bibfield  {author} {\bibinfo {author} {\bibfnamefont {ZZ}~\bibnamefont
  {Du}}, \bibinfo {author} {\bibfnamefont {CM}~\bibnamefont {Wang}}, \bibinfo
  {author} {\bibfnamefont {Shuai}\ \bibnamefont {Li}}, \bibinfo {author}
  {\bibfnamefont {Hai-Zhou}\ \bibnamefont {Lu}}, \ and\ \bibinfo {author}
  {\bibfnamefont {XC}~\bibnamefont {Xie}},\ }\bibfield  {title} {\enquote
  {\bibinfo {title} {Disorder-induced nonlinear hall effect with time-reversal
  symmetry},}\ }\href@noop {} {\bibfield  {journal} {\bibinfo  {journal}
  {Nature Communications}\ }\textbf {\bibinfo {volume} {10}},\ \bibinfo {pages}
  {3047} (\bibinfo {year} {2019})}\BibitemShut {NoStop}%
\bibitem [{\citenamefont {Isobe}\ \emph {et~al.}(2020)\citenamefont {Isobe},
  \citenamefont {Xu},\ and\ \citenamefont {Fu}}]{Isobeeaay2497}%
  \BibitemOpen
  \bibfield  {author} {\bibinfo {author} {\bibfnamefont {Hiroki}\ \bibnamefont
  {Isobe}}, \bibinfo {author} {\bibfnamefont {Su-Yang}\ \bibnamefont {Xu}}, \
  and\ \bibinfo {author} {\bibfnamefont {Liang}\ \bibnamefont {Fu}},\
  }\bibfield  {title} {\enquote {\bibinfo {title} {High-frequency rectification
  via chiral bloch electrons},}\ }\href {\doibase 10.1126/sciadv.aay2497}
  {\bibfield  {journal} {\bibinfo  {journal} {Science Advances}\ }\textbf
  {\bibinfo {volume} {6}} (\bibinfo {year} {2020}),\
  10.1126/sciadv.aay2497}\BibitemShut {NoStop}%
\bibitem [{\citenamefont {Song}\ and\ \citenamefont
  {Dai}(2019)}]{PhysRevX.9.021053}%
  \BibitemOpen
  \bibfield  {author} {\bibinfo {author} {\bibfnamefont {Zhida}\ \bibnamefont
  {Song}}\ and\ \bibinfo {author} {\bibfnamefont {Xi}~\bibnamefont {Dai}},\
  }\bibfield  {title} {\enquote {\bibinfo {title} {Hear the sound of weyl
  fermions},}\ }\href {\doibase 10.1103/PhysRevX.9.021053} {\bibfield
  {journal} {\bibinfo  {journal} {Phys. Rev. X}\ }\textbf {\bibinfo {volume}
  {9}},\ \bibinfo {pages} {021053} (\bibinfo {year} {2019})}\BibitemShut
  {NoStop}%
\bibitem [{\citenamefont {Xiang}\ \emph {et~al.}(2019)\citenamefont {Xiang},
  \citenamefont {Hu}, \citenamefont {Song}, \citenamefont {Lv}, \citenamefont
  {Zhang}, \citenamefont {Zhao}, \citenamefont {Li}, \citenamefont {Chen},
  \citenamefont {Zhang}, \citenamefont {Wang}, \citenamefont {Yang},
  \citenamefont {Dai}, \citenamefont {Steglich}, \citenamefont {Chen},\ and\
  \citenamefont {Sun}}]{PhysRevX.9.031036}%
  \BibitemOpen
  \bibfield  {author} {\bibinfo {author} {\bibfnamefont {Junsen}\ \bibnamefont
  {Xiang}}, \bibinfo {author} {\bibfnamefont {Sile}\ \bibnamefont {Hu}},
  \bibinfo {author} {\bibfnamefont {Zhida}\ \bibnamefont {Song}}, \bibinfo
  {author} {\bibfnamefont {Meng}\ \bibnamefont {Lv}}, \bibinfo {author}
  {\bibfnamefont {Jiahao}\ \bibnamefont {Zhang}}, \bibinfo {author}
  {\bibfnamefont {Lingxiao}\ \bibnamefont {Zhao}}, \bibinfo {author}
  {\bibfnamefont {Wei}\ \bibnamefont {Li}}, \bibinfo {author} {\bibfnamefont
  {Ziyu}\ \bibnamefont {Chen}}, \bibinfo {author} {\bibfnamefont {Shuai}\
  \bibnamefont {Zhang}}, \bibinfo {author} {\bibfnamefont {Jian-Tao}\
  \bibnamefont {Wang}}, \bibinfo {author} {\bibfnamefont {Yi-feng}\
  \bibnamefont {Yang}}, \bibinfo {author} {\bibfnamefont {Xi}~\bibnamefont
  {Dai}}, \bibinfo {author} {\bibfnamefont {Frank}\ \bibnamefont {Steglich}},
  \bibinfo {author} {\bibfnamefont {Genfu}\ \bibnamefont {Chen}}, \ and\
  \bibinfo {author} {\bibfnamefont {Peijie}\ \bibnamefont {Sun}},\ }\bibfield
  {title} {\enquote {\bibinfo {title} {Giant magnetic quantum oscillations in
  the thermal conductivity of taas: Indications of chiral zero sound},}\ }\href
  {\doibase 10.1103/PhysRevX.9.031036} {\bibfield  {journal} {\bibinfo
  {journal} {Phys. Rev. X}\ }\textbf {\bibinfo {volume} {9}},\ \bibinfo {pages}
  {031036} (\bibinfo {year} {2019})}\BibitemShut {NoStop}%
\bibitem [{\citenamefont {Coulter}\ \emph {et~al.}(2019)\citenamefont
  {Coulter}, \citenamefont {Osterhoudt}, \citenamefont {Garcia}, \citenamefont
  {Wang}, \citenamefont {Plisson}, \citenamefont {Shen}, \citenamefont {Ni},
  \citenamefont {Burch},\ and\ \citenamefont {Narang}}]{PhysRevB.100.220301}%
  \BibitemOpen
  \bibfield  {author} {\bibinfo {author} {\bibfnamefont {Jennifer}\
  \bibnamefont {Coulter}}, \bibinfo {author} {\bibfnamefont {Gavin~B.}\
  \bibnamefont {Osterhoudt}}, \bibinfo {author} {\bibfnamefont {Christina
  A.~C.}\ \bibnamefont {Garcia}}, \bibinfo {author} {\bibfnamefont {Yiping}\
  \bibnamefont {Wang}}, \bibinfo {author} {\bibfnamefont {Vincent~M.}\
  \bibnamefont {Plisson}}, \bibinfo {author} {\bibfnamefont {Bing}\
  \bibnamefont {Shen}}, \bibinfo {author} {\bibfnamefont {Ni}~\bibnamefont
  {Ni}}, \bibinfo {author} {\bibfnamefont {Kenneth~S.}\ \bibnamefont {Burch}},
  \ and\ \bibinfo {author} {\bibfnamefont {Prineha}\ \bibnamefont {Narang}},\
  }\bibfield  {title} {\enquote {\bibinfo {title} {Uncovering electron-phonon
  scattering and phonon dynamics in type-i weyl semimetals},}\ }\href {\doibase
  10.1103/PhysRevB.100.220301} {\bibfield  {journal} {\bibinfo  {journal}
  {Phys. Rev. B}\ }\textbf {\bibinfo {volume} {100}},\ \bibinfo {pages}
  {220301(R)} (\bibinfo {year} {2019})}\BibitemShut {NoStop}%
\bibitem [{\citenamefont {Zhang}\ and\ \citenamefont
  {Zhou}(2020)}]{PhysRevB.101.085202}%
  \BibitemOpen
  \bibfield  {author} {\bibinfo {author} {\bibfnamefont {Song-Bo}\ \bibnamefont
  {Zhang}}\ and\ \bibinfo {author} {\bibfnamefont {Jianhui}\ \bibnamefont
  {Zhou}},\ }\bibfield  {title} {\enquote {\bibinfo {title} {Quantum
  oscillations in acoustic phonons in weyl semimetals},}\ }\href {\doibase
  10.1103/PhysRevB.101.085202} {\bibfield  {journal} {\bibinfo  {journal}
  {Phys. Rev. B}\ }\textbf {\bibinfo {volume} {101}},\ \bibinfo {pages}
  {085202} (\bibinfo {year} {2020})}\BibitemShut {NoStop}%
\bibitem [{\citenamefont {Lalibert\'e}\ \emph {et~al.}(2020)\citenamefont
  {Lalibert\'e}, \citenamefont {B\'elanger}, \citenamefont {Nair},
  \citenamefont {Analytis}, \citenamefont {Boulanger}, \citenamefont {Dion},
  \citenamefont {Taillefer},\ and\ \citenamefont
  {Quilliam}}]{PhysRevB.102.125104}%
  \BibitemOpen
  \bibfield  {author} {\bibinfo {author} {\bibfnamefont {F.}~\bibnamefont
  {Lalibert\'e}}, \bibinfo {author} {\bibfnamefont {F.}~\bibnamefont
  {B\'elanger}}, \bibinfo {author} {\bibfnamefont {N.~L.}\ \bibnamefont
  {Nair}}, \bibinfo {author} {\bibfnamefont {J.~G.}\ \bibnamefont {Analytis}},
  \bibinfo {author} {\bibfnamefont {M.-E.}\ \bibnamefont {Boulanger}}, \bibinfo
  {author} {\bibfnamefont {M.}~\bibnamefont {Dion}}, \bibinfo {author}
  {\bibfnamefont {L.}~\bibnamefont {Taillefer}}, \ and\ \bibinfo {author}
  {\bibfnamefont {J.~A.}\ \bibnamefont {Quilliam}},\ }\bibfield  {title}
  {\enquote {\bibinfo {title} {Field-angle dependence of sound velocity in the
  weyl semimetal taas},}\ }\href {\doibase 10.1103/PhysRevB.102.125104}
  {\bibfield  {journal} {\bibinfo  {journal} {Phys. Rev. B}\ }\textbf {\bibinfo
  {volume} {102}},\ \bibinfo {pages} {125104} (\bibinfo {year}
  {2020})}\BibitemShut {NoStop}%
\bibitem [{\citenamefont {Sengupta}\ \emph {et~al.}(2020)\citenamefont
  {Sengupta}, \citenamefont {Lhachemi},\ and\ \citenamefont
  {Garate}}]{PhysRevLett.125.146402}%
  \BibitemOpen
  \bibfield  {author} {\bibinfo {author} {\bibfnamefont {Sanghita}\
  \bibnamefont {Sengupta}}, \bibinfo {author} {\bibfnamefont {M.~Nabil~Y.}\
  \bibnamefont {Lhachemi}}, \ and\ \bibinfo {author} {\bibfnamefont {Ion}\
  \bibnamefont {Garate}},\ }\bibfield  {title} {\enquote {\bibinfo {title}
  {Phonon magnetochiral effect of band-geometric origin in weyl semimetals},}\
  }\href {\doibase 10.1103/PhysRevLett.125.146402} {\bibfield  {journal}
  {\bibinfo  {journal} {Phys. Rev. Lett.}\ }\textbf {\bibinfo {volume} {125}},\
  \bibinfo {pages} {146402} (\bibinfo {year} {2020})}\BibitemShut {NoStop}%
\bibitem [{\citenamefont {Inoue}\ \emph {et~al.}(2016)\citenamefont {Inoue},
  \citenamefont {Gyenis}, \citenamefont {Wang}, \citenamefont {Li},
  \citenamefont {Oh}, \citenamefont {Jiang}, \citenamefont {Ni}, \citenamefont
  {Bernevig},\ and\ \citenamefont {Yazdani}}]{Inoue1184}%
  \BibitemOpen
  \bibfield  {author} {\bibinfo {author} {\bibfnamefont {Hiroyuki}\
  \bibnamefont {Inoue}}, \bibinfo {author} {\bibfnamefont {Andr{\'a}s}\
  \bibnamefont {Gyenis}}, \bibinfo {author} {\bibfnamefont {Zhijun}\
  \bibnamefont {Wang}}, \bibinfo {author} {\bibfnamefont {Jian}\ \bibnamefont
  {Li}}, \bibinfo {author} {\bibfnamefont {Seong~Woo}\ \bibnamefont {Oh}},
  \bibinfo {author} {\bibfnamefont {Shan}\ \bibnamefont {Jiang}}, \bibinfo
  {author} {\bibfnamefont {Ni}~\bibnamefont {Ni}}, \bibinfo {author}
  {\bibfnamefont {B.~Andrei}\ \bibnamefont {Bernevig}}, \ and\ \bibinfo
  {author} {\bibfnamefont {Ali}\ \bibnamefont {Yazdani}},\ }\bibfield  {title}
  {\enquote {\bibinfo {title} {Quasiparticle interference of the fermi arcs and
  surface-bulk connectivity of a weyl semimetal},}\ }\href {\doibase
  10.1126/science.aad8766} {\bibfield  {journal} {\bibinfo  {journal}
  {Science}\ }\textbf {\bibinfo {volume} {351}},\ \bibinfo {pages} {1184--1187}
  (\bibinfo {year} {2016})}\BibitemShut {NoStop}%
\bibitem [{\citenamefont {Xu}\ \emph {et~al.}(2015)\citenamefont {Xu},
  \citenamefont {Alidoust}, \citenamefont {Belopolski}, \citenamefont {Yuan},
  \citenamefont {Bian}, \citenamefont {Chang}, \citenamefont {Zheng},
  \citenamefont {Strocov}, \citenamefont {Sanchez}, \citenamefont {Chang} \emph
  {et~al.}}]{xu2015discovery}%
  \BibitemOpen
  \bibfield  {author} {\bibinfo {author} {\bibfnamefont {Su-Yang}\ \bibnamefont
  {Xu}}, \bibinfo {author} {\bibfnamefont {Nasser}\ \bibnamefont {Alidoust}},
  \bibinfo {author} {\bibfnamefont {Ilya}\ \bibnamefont {Belopolski}}, \bibinfo
  {author} {\bibfnamefont {Zhujun}\ \bibnamefont {Yuan}}, \bibinfo {author}
  {\bibfnamefont {Guang}\ \bibnamefont {Bian}}, \bibinfo {author}
  {\bibfnamefont {Tay-Rong}\ \bibnamefont {Chang}}, \bibinfo {author}
  {\bibfnamefont {Hao}\ \bibnamefont {Zheng}}, \bibinfo {author} {\bibfnamefont
  {Vladimir~N}\ \bibnamefont {Strocov}}, \bibinfo {author} {\bibfnamefont
  {Daniel~S}\ \bibnamefont {Sanchez}}, \bibinfo {author} {\bibfnamefont
  {Guoqing}\ \bibnamefont {Chang}},  \emph {et~al.},\ }\bibfield  {title}
  {\enquote {\bibinfo {title} {Discovery of a weyl fermion state with fermi
  arcs in niobium arsenide},}\ }\href@noop {} {\bibfield  {journal} {\bibinfo
  {journal} {Nature Physics}\ }\textbf {\bibinfo {volume} {11}},\ \bibinfo
  {pages} {748--754} (\bibinfo {year} {2015})}\BibitemShut {NoStop}%
\bibitem [{\citenamefont {Bruno}\ \emph {et~al.}(2016)\citenamefont {Bruno},
  \citenamefont {Tamai}, \citenamefont {Wu}, \citenamefont {Cucchi},
  \citenamefont {Barreteau}, \citenamefont {de~la Torre}, \citenamefont
  {McKeown~Walker}, \citenamefont {Ricc\`o}, \citenamefont {Wang},
  \citenamefont {Kim}, \citenamefont {Hoesch}, \citenamefont {Shi},
  \citenamefont {Plumb}, \citenamefont {Giannini}, \citenamefont {Soluyanov},\
  and\ \citenamefont {Baumberger}}]{PhysRevB.94.121112}%
  \BibitemOpen
  \bibfield  {author} {\bibinfo {author} {\bibfnamefont {F.~Y.}\ \bibnamefont
  {Bruno}}, \bibinfo {author} {\bibfnamefont {A.}~\bibnamefont {Tamai}},
  \bibinfo {author} {\bibfnamefont {Q.~S.}\ \bibnamefont {Wu}}, \bibinfo
  {author} {\bibfnamefont {I.}~\bibnamefont {Cucchi}}, \bibinfo {author}
  {\bibfnamefont {C.}~\bibnamefont {Barreteau}}, \bibinfo {author}
  {\bibfnamefont {A.}~\bibnamefont {de~la Torre}}, \bibinfo {author}
  {\bibfnamefont {S.}~\bibnamefont {McKeown~Walker}}, \bibinfo {author}
  {\bibfnamefont {S.}~\bibnamefont {Ricc\`o}}, \bibinfo {author} {\bibfnamefont
  {Z.}~\bibnamefont {Wang}}, \bibinfo {author} {\bibfnamefont {T.~K.}\
  \bibnamefont {Kim}}, \bibinfo {author} {\bibfnamefont {M.}~\bibnamefont
  {Hoesch}}, \bibinfo {author} {\bibfnamefont {M.}~\bibnamefont {Shi}},
  \bibinfo {author} {\bibfnamefont {N.~C.}\ \bibnamefont {Plumb}}, \bibinfo
  {author} {\bibfnamefont {E.}~\bibnamefont {Giannini}}, \bibinfo {author}
  {\bibfnamefont {A.~A.}\ \bibnamefont {Soluyanov}}, \ and\ \bibinfo {author}
  {\bibfnamefont {F.}~\bibnamefont {Baumberger}},\ }\bibfield  {title}
  {\enquote {\bibinfo {title} {Observation of large topologically trivial fermi
  arcs in the candidate type-ii weyl semimetal
  $\mathrm{WT}{\mathrm{e}}_{2}$},}\ }\href {\doibase
  10.1103/PhysRevB.94.121112} {\bibfield  {journal} {\bibinfo  {journal} {Phys.
  Rev. B}\ }\textbf {\bibinfo {volume} {94}},\ \bibinfo {pages} {121112(R)}
  (\bibinfo {year} {2016})}\BibitemShut {NoStop}%
\bibitem [{\citenamefont {Bradlyn}\ \emph {et~al.}(2016)\citenamefont
  {Bradlyn}, \citenamefont {Cano}, \citenamefont {Wang}, \citenamefont
  {Vergniory}, \citenamefont {Felser}, \citenamefont {Cava},\ and\
  \citenamefont {Bernevig}}]{Bradlynaaf5037}%
  \BibitemOpen
  \bibfield  {author} {\bibinfo {author} {\bibfnamefont {Barry}\ \bibnamefont
  {Bradlyn}}, \bibinfo {author} {\bibfnamefont {Jennifer}\ \bibnamefont
  {Cano}}, \bibinfo {author} {\bibfnamefont {Zhijun}\ \bibnamefont {Wang}},
  \bibinfo {author} {\bibfnamefont {M.~G.}\ \bibnamefont {Vergniory}}, \bibinfo
  {author} {\bibfnamefont {C.}~\bibnamefont {Felser}}, \bibinfo {author}
  {\bibfnamefont {R.~J.}\ \bibnamefont {Cava}}, \ and\ \bibinfo {author}
  {\bibfnamefont {B.~Andrei}\ \bibnamefont {Bernevig}},\ }\bibfield  {title}
  {\enquote {\bibinfo {title} {Beyond dirac and weyl fermions: Unconventional
  quasiparticles in conventional crystals},}\ }\href {\doibase
  10.1126/science.aaf5037} {\bibfield  {journal} {\bibinfo  {journal}
  {Science}\ }\textbf {\bibinfo {volume} {353}} (\bibinfo {year} {2016}),\
  10.1126/science.aaf5037}\BibitemShut {NoStop}%
\end{thebibliography}%

\end{document}